\documentclass[11pt]{article}

\pdfoutput=1

\usepackage{booktabs} 

\usepackage{float}
\usepackage{tikz-feynman}
\tikzfeynmanset{compat=1.1.0}
\usepackage{caption,subcaption}
\usepackage{multirow}

\pdfoutput=1
\usepackage{latexsym}
\usepackage{epsfig}
\usepackage[mathscr]{eucal}
\usepackage{amsfonts}
\usepackage{amscd}
\usepackage{array}
\usepackage{bbold}
\usepackage{amssymb}
\usepackage{colordvi}
\usepackage[centertags]{amsmath}
\usepackage{enumerate}
\usepackage{graphicx}
\usepackage{booktabs}
\usepackage{theorem}
\usepackage{soul}
\usepackage[numbers]{natbib}
\usepackage{url}
\usepackage[final]{hyperref} 
\hypersetup{
	colorlinks=true,    
	linktocpage=true,   
    linkcolor=blue,     
    citecolor=green,    
    filecolor=cyan,     
    urlcolor=magenta    
}
\usepackage{slashed}
\usepackage[normalem]{ulem}
\usepackage{bbm}
\usepackage[utf8]{inputenc}
\usepackage{braket}

\allowdisplaybreaks{}

\numberwithin{equation}{section}

\newcommand{\be}{\begin{equation}}
\newcommand{\ee}{\end{equation}}
\newcommand{\beq}{\begin{eqnarray}}
\newcommand{\eeq}{\end{eqnarray}}

\newcommand{\RNum}[1]{\uppercase\expandafter{\romannumeral #1\relax}}

\newcommand{\si}[1]{\text}
\newcommand{\SI}[1]{\text}

\usepackage{comment}

\begin{document}

\title{\textbf{Dark Matter in Multi-Singlet Extensions of the Standard Model}}

\date{\today}
\author{
Maria Gonçalves$^{1,2\,}$\footnote{E-mail: \href{mailto:magoncalves@fc.ul.pt}{\texttt{magoncalves@fc.ul.pt}}},
Margarete M\"{u}hlleitner$^{2\,}$\footnote{E-mail:
	\href{mailto:margarete.muehlleitner@kit.edu}{\texttt{margarete.muehlleitner@kit.edu}}},
Rui Santos$^{1,3\,}$\footnote{E-mail:  \href{mailto:rasantos@fc.ul.pt}{\texttt{rasantos@fc.ul.pt}}},
Tomás Trindade$^{1\,}$\footnote{E-mail:  \href{mailto:ttrindadehenriques@gmail.com}{\texttt{ttrindadehenriques@gmail.com}}}
\\[9mm]
{\small\it
$^1$Centro de F\'{\i}sica Te\'{o}rica e Computacional,
    Faculdade de Ci\^{e}ncias,} \\
{\small \it    Universidade de Lisboa, Campo Grande, Edif\'{\i}cio C8
  1749-016 Lisboa, Portugal} \\[3mm]
{\small\it
$^2$Institute for Theoretical Physics, Karlsruhe Institute of Technology,} \\
{\small\it Wolfgang-Gaede-Str. 1, 76131 Karlsruhe, Germany.}\\[3mm]
{\small\it
$^3$ISEL -
 Instituto Superior de Engenharia de Lisboa,} \\
{\small \it   Instituto Polit\'ecnico de Lisboa
 1959-007 Lisboa, Portugal} \\[3mm]
}

\maketitle

\begin{abstract}
\noindent
We study the simplest extensions of the Standard Model (SM) that provide Dark Matter (DM) candidates, built with the addition of real singlets and new $\mathcal{Z}_2$ symmetries.
In this type of models the interactions between SM particles are not altered except for the new interactions stemming from the portal couplings that link the SM Higgs with the DM candidates. In the extension with just one singlet, DM masses below about 3.5 TeV are already excluded by the combination of relic density and direct detection (DD) constraints, except in the resonant case where the DM mass is close to half the Higgs mass, making them undetectable at the LHC. Adding just one more real singlet with an independent $\mathcal{Z}_2$ symmetry opens up a new mass window for one of the DM candidates and decreases the lower bound on the mass of the other. 
Adding more singlets with independent $\mathcal{Z}_2$ symmetries will not change this picture dramatically. If instead we add new singlets all odd under the same $\mathcal{Z}_2$ symmetry, the allowed mass region for the  DM candidate (i.e., the lightest dark sector scalar) will span the entire mass range from half the Higgs mass to the TeV scale.  In principle, such light particles could be probed at the LHC in mono-$X$  searches. Although they are still out of reach with the current LHC DM searches, there are good chances to probe the models in some final states at the High-Luminosity (HL-LHC) stage of the LHC.

\end{abstract}

\thispagestyle{empty}
\vfill
\newpage
\setcounter{page}{1}

\section{Introduction}

Although the existence of Dark Matter (DM) was first mentioned about 100 years ago~\cite{Zwicky:1933gu} we still do not know if it can be explained by the introduction of a new field in some extension of the Standard Model (SM). 
This new field is usually considered to live in a dark sector that connects with the visible world via a portal term in the Lagrangian~\cite{Patt:2006fw}.
The dark sector camouflage is a symmetry under which the visible fields are even while the ones from the dark sector are odd. In this work we will use  $\mathcal{Z}_2$  symmetries to disconnect the two sectors, but any symmetry that stabilises the DM candidate accomplishes the goal.
Once the SM is extended to accommodate one or more DM candidates, there are several experimental constraints that curtail the parameter space of the model. We will consider DM particles that are produced via the so-called freeze-out~\cite{Zeldovich:1965gev, Bertone:2004pz, Feng:2010gw, Cirelli:2024ssz} mechanism. 
This is the scenario where the previous history of the universe is forgotten. The DM candidates are in thermal equilibrium with the thermal bath and DM annihilation ends when the rate of expansion of the universe becomes larger than the annihilation rate into lighter particles. Therefore, we consider that our DM candidates are Weakly Interacting Massive Particles (WIMPs). 

One of the simplest ways of extending the SM to include a DM particle, without breaking the SM gauge symmetry or spoil its renormalisability, is by adding a real scalar singlet~\cite{Silveira:1985rk, Burgess:2000yq, Barger:2007im, Guo:2010hq} that only couples to the Higgs doublet. This minimal model, with an unbroken $\mathcal{Z}_2$  symmetry provides a DM particle while keeping all other SM couplings unchanged. The only way to detect the new particle is via its interactions with the Higgs boson originating from the portal coupling. However, this particular model is heavily constrained by experiment -- in particular, by the relic density measurement and by direct detection (DD) experiments. Although other constraints like indirect detection are also relevant, for the mass window we will be exploring, above about 125 GeV,  these are the two relevant constraints. The searches for DM at the Large Hadron Collider (LHC) will also be discussed in detail.

For this extension of the SM and for the mass range considered, the allowed parameter space results from a tension between DD constraints and the relic density measured by PLANCK~\cite{Planck:2018vyg}. For a given mass, DD enforces an upper bound on the coupling while the relic density enforces a lower bound on the same portal coupling, except for the scenario of resonance in DM annihilation cross sections (when the DM mass is close to half of the Higgs boson mass). As the mass grows, the experimental DD bounds get weaker and a DM mass above about 3.5 TeV together with a portal coupling above 1 are allowed. DM masses above 3.5 TeV will not be probed at the LHC due to the negligible values of the DM production cross section, even for the maximally allowed coupling. There is also an allowed  region around a DM mass of half the Higgs boson mass. This region is allowed for a small portal coupling because of the resonant cross section for DM production. This way the enhancement of the cross section is not due to a large coupling which allows for an agreement with the DD bounds.

In this work we analyse the addition of more singlets to the minimal model with just one singlet. These extensions will still leave the SM interactions unchanged. They can either add more DM candidates, if for each new scalar a new $\mathcal{Z}_2$ symmetry is imposed, or have just one DM particle if only one $\mathcal{Z}_2$ symmetry is imposed. The question we want to answer now is whether such extensions will lead to different phenomenology and the opening of allowed parameter space, with mass ranges that could be probed at the LHC or at future colliders. 

The paper is organised as follows.
 Sec.~\ref{sec:2} is devoted to the two real singlets extension of the SM and Sec.~\ref{sec:3} to the three real singlets extension, based on two and three independent $\mathcal{Z}_2 $ symmetries, respectively. For each extension, we first introduce the model and then present the parameter scan and numerical analysis. For the two real singlets extension we also discuss the case with only one independent $\mathcal{Z}_2$ symmetry. In Sec.~\ref{sec:4}, we discuss DM searches at the LHC. Our conclusions are given in Sec.~\ref{sec:conclusions}.

\section{Two Real Singlets Extension of the SM}
\label{sec:2}

DM models with two real singlets have been studied in the context of various symmetries and DM production mechanisms~\cite{Bhattacharya:2016ysw,Bhattacharya:2017fid,Maity:2019hre,Bhattacharya:2024nla,Capucha:2024oaa}. Here, we focus on the $\mathcal{Z}_2^{(1)} \times \mathcal{Z}_2^{(2)}$ and $\mathcal{Z}_2$ symmetric models, assuming freeze-out as the production mechanism for all DM candidates, and take into account the latest experimental constraints.

\subsection{Two Independent $\mathcal{Z}_2$ Symmetries}

\begin{figure}[h]
    \centering
    \includegraphics[width=0.95\textwidth]{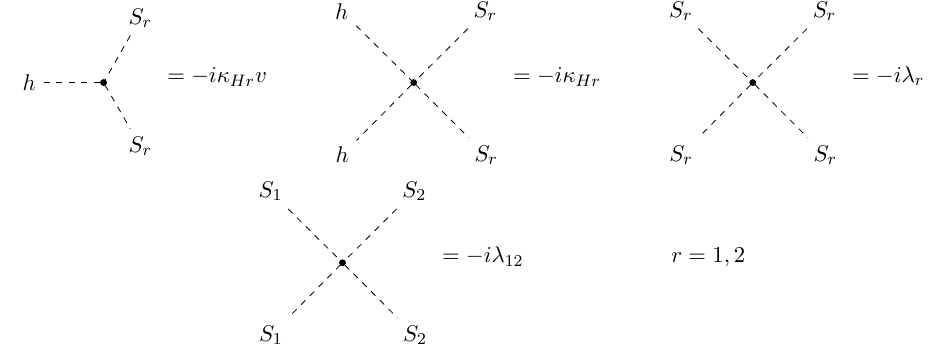}
    \caption{Feynman rules for the vertices involving the $S_1$ and $S_2$ DM particles.}
    \label{fig:SM+2RSS vertices}
\end{figure}

The model discussed in this section is an extension of the SM with two real scalar singlet fields $S_1, S_2 \sim (\mathbf{1}, \mathbf{1}, 0)$,\footnote{This notation refers to the $\mathcal{G}_{\text{SM}} \equiv \text{SU}(3)_c \times \text{SU}(2)_L \times \text{U}(1)_Y$ representation under which a field transforms.} with a Lagrangian invariant under $\mathcal{Z}^{(1)}_2 \times \mathcal{Z}^{(2)}_2$ transformations,
\begin{align}
    \mathcal{Z}^{(1)}_2 : \quad &S_1 \rightarrow - S_1, \quad S_2 \rightarrow + S_2, \quad \text{SM} \rightarrow +\text{SM}\\
    \mathcal{Z}^{(2)}_2 : \quad &S_1 \rightarrow + S_1, \quad S_2 \rightarrow - S_2, \quad \text{SM} \rightarrow +\text{SM}\, .
\end{align}
The $\mathcal{Z}^{(r)}_2$ ($r=1,2$) charges are two independent \emph{dark}  parity quantum numbers. Therefore, the most general renormalisable and $\text{SU}(3)_c \times \text{SU}(2)_L \times \text{U}(1)_Y \times \mathcal{Z}_2^{(1)} \times \mathcal{Z}_2^{(2)}$ invariant Lagrangian is given by
\begin{align}
    \notag \mathcal{L}_{\text{SM+2RSS}} \, = \: &\mathcal{L}_{\text{SM}} + \frac{1}{2} (\partial_{\mu} S_1) \partial^{\mu}S_1 - \frac{1}{2}\mu_1^2 S_1^2 + \frac{1}{2} (\partial_{\mu} S_2) \partial^{\mu}S_2 - \frac{1}{2} \mu_2^2 S_2^2 - \frac{\lambda_1}{4!} S_1^4 - \frac{\lambda_2}{4!} S_2^4 \\
    &\underbrace{- \frac{\kappa_{H1}}{2}S_1^2 \Phi^{\dagger} \Phi}_{= \, \mathcal{L}_{\text{portal}(1)} }
    \underbrace{- \frac{\kappa_{H2}}{2}S_2^2 \Phi^{\dagger} \Phi}_{= \, \mathcal{L}_{\text{portal}(2)} } \underbrace{- \frac{\lambda_{12}}{4} S_1^2 S_2^2}_{= \, \mathcal{L}_{\text{int}(1,2)}} \, ,
    \label{eq:SM+2RSS}
\end{align}
where $\Phi = \begin{pmatrix} G^+, & \phi^0 \end{pmatrix}^{\text{T}} \sim (\mathbf{1}, \mathbf{2}, +1/2)$ is the Higgs doublet. The vacuum state is obtained by the usual minimisation procedure, and from the eight $\text{SU}(3)_c \times \text{SU}(2)_L \times \text{U}(1)_Y \times \mathcal{Z}^{(1)}_2 \times \mathcal{Z}^{(2)}_2$ invariant solution sets,\footnote{The analysis of the different minima and the possibility of tunneling is beyond the scope of this work.} we consider
\begin{equation}
\braket{\Phi^\dagger \Phi}_0 = -\frac{\mu_H^2}{2\lambda_H} \equiv \frac{v^2}{2} \, , \quad \braket{S_r}_0 = 0 \, , \quad r=1,2  \label{eq:SM+2RSS vacuum 2}
\end{equation}
as the vacuum configuration of the model and take $\Phi = \begin{pmatrix} G^+, & ( v + h + i G^0)/\sqrt{2} \end{pmatrix}^{\text{T}}$, where $h$ is the Higgs boson and $G^0, G^\pm$ are the would-be Goldstone bosons. For this minimum solution set, the two singlets do not acquire a vacuum expectation value (VEV), so that the $\mathcal{Z}^{(1)}_2 \times \mathcal{Z}^{(2)}_2$ symmetry is not spontaneously broken and there is no mixing with the Higgs.\footnote{Scenarios where $\mathcal{Z}_2$ symmetries are spontaneously broken were discussed e.g. in~\cite{Barger:2008jx, Costa:2015llh} for the complex singlet extension of the SM and in~\cite{Robens:2019kga} for the two real singlets extension of the SM.} Using the minimum condition $v^2=-\mu_H^2/\lambda_H$ in~\eqref{eq:SM+2RSS vacuum 2}, the scalar potential can be written in the unitary gauge $\Phi = 1/\sqrt{2} \begin{pmatrix} 0, & v + h \end{pmatrix}^{\text{T}}$ as
\begin{align}
    \notag V(|\Phi|, S_{r=1,2}) \, = \, &\mu_H^2 \Phi^{\dagger}\Phi + \lambda_H ( \Phi^{\dagger}\Phi )^2  + \sum_{r=1}^{2} \left[ \frac{1}{2}\mu_r^2 S_r^2 + \frac{\lambda_{r}}{4!}S_r^4 + \frac{\kappa_{Hr}}{2}S_r^2 \Phi^{\dagger} \Phi \right] + \frac{\lambda_{12}}{4}S_1^2 S_2^2 \\
    \notag = \, &\frac{1}{2}\overbrace{(2 \lambda_H v^2)}^{= \, m_h^2} h^2 + \lambda_H v h^3 + \frac{\lambda_H}{4}h^4 + \frac{\lambda_{12}}{4}S_1^2 S_2^2\\ &+ \sum_{r=1}^{2} \bigg[ \frac{1}{2}\underbrace{(\mu_r^2 + \frac{\kappa_{Hr}v^2}{2})}_{= \, m^2_{S_r}} S_r^2 + \frac{\lambda_{r}}{4!}S_r^4
    + \frac{\kappa_{Hr}v}{2} h S_r^2 + \frac{\kappa_{Hr}}{4} h^2 S_r^2 \bigg] \, .
\end{align}
The unbroken $\mathcal{Z}^{(1)}_2 \times \mathcal{Z}^{(2)}_2$ symmetry ensures that $S_r$ ($r=1,2$) do not decay, thus being DM candidates. This is shown in Fig.~\ref{fig:SM+2RSS vertices}, which presents the new Feynman rules for this model.

\subsubsection{Theoretical and Experimental Constraints}

\textbf{Experimental constraints:} this model is not affected by the LHC results on cross sections and branching ratios involving SM particles – these quantities were calculated at leading order (LO) in this work. We note that in order to probe the model we have generated events using \texttt{MadGraph5\_aMC@NLO}~\cite{Alwall:2011uj, Alwall:2014hca} and performed the cuts presented in the experimental analysis. While the total Higgs production cross sections at LO can be off by 50\%, with a set of specific cuts this number can change. However, as will become clear when we present the results, this factor will not play a role in the exclusion since the cross sections are still far away from the experimental bounds.
In fact, since so far all results are in agreement with the SM predictions, there are 
no constraints on the model from the Higgs data and new Higgs searches. The model can  only be tested by searches for DM including mono-jet, mono-Higgs, mono-$Z$ events, among others. The discussion
on these searches will be presented later. Since $\Gamma(h \rightarrow S_r S_r) = \kappa_{Hr}^2 v^2 /(32 \pi m_h) \sqrt{1- 4m_{S_r}^2/m_h^2}$ ($r=1,2$), the only LHC constraint from Higgs physics we have to take into account is the upper limit on the invisible Higgs branching ratio, $\text{BR}(h \rightarrow \text{inv})<0.107$~\cite{ATLAS:2023tkt}.
Since the $\mathcal{Z}_2$ symmetries remain exact there are also no new contributions to the gauge bosons self-energies and therefore to the $S$, $T$ and $U$ parameters~\cite{Peskin:1991sw}.

\vspace{0.5em}
\noindent \textbf{Theoretical constraints:} We take all quartic parameters to be below $4 \pi$. The boundedness-from-below conditions, derived from copositivity criteria~\cite{Kannike:2012pe, Kannike:2016fmd} and presented in~\cite{Robens:2019kga}, take the following form in our notation:
\begin{align}
	&\lambda_H, \lambda_r > 0 \, , \quad r=1,2 \, ,\\
	&\underline{\kappa}_{Hr}/2 \equiv \kappa_{Hr}/2 + \sqrt{\lambda_H \lambda_r/6} > 0 \, , \quad r=1,2 \, ,\\
	&\underline{\lambda}_{12} \equiv \lambda_{12} + \sqrt{\lambda_1 \lambda_2} /3 > 0 \, , \\
	&\kappa_{H1} \sqrt{\lambda_2 /6} + \kappa_{H2} \sqrt{\lambda_1 /6} + \lambda_{12}\sqrt{\lambda_H} + \sqrt{\lambda_H \lambda_1 \lambda_2}/6 + \sqrt{\underline{\kappa}_{H1} \underline{\kappa}_{H2} \underline{\lambda}_{12}} > 0 \, .
\end{align}
The conditions for tree-level perturbative unitarity can be obtained by imposing $|\text{Re}(a_0)|<1/2$ for all the relevant $2\to2$ scalar processes in the high-energy regime $s\gg m_{h}^2, m_{S_r}^2$, where
\begin{equation}
	a_0(s) = \frac{1}{32\pi} \int_{-1}^{+1} P_l(\cos{\theta}) \mathcal{M}(s,\cos{\theta}) \, d\cos{\theta} \bigg|_{l=0} = \frac{1}{32\pi} \int_{-1}^{+1} \mathcal{M}(s,\cos{\theta}) \, d\cos{\theta} \label{eq:a_0}
\end{equation}
is the coefficient $a_l(s)$ of order $l=0$ of the partial-wave expansion (in Legendre polynomials $P_l$) of the transition amplitude $\mathcal{M}(s,\cos{\theta}) = 16\pi \sum_{l=0}^\infty (2l+1)P_l(\cos{\theta}) a_l(s)$, and are given by
\begin{align}
	&16\pi \big| a_0(hh \rightarrow hh) \big|_{s\gg m_{h}^2, m_{S_r}^2} = \left| 3m_h^2/v^2 \right| < 8\pi \, ,\label{eq:PU1}\\
	&16\pi \big| a_0(S_r S_r \rightarrow S_r S_r) \big|_{s\gg m_{h}^2, m_{S_r}^2} = | \lambda_r | < 8\pi \, ,\\
	&16\pi \big| a_0(i \rightarrow f) \big|_{s\gg m_{h}^2, m_{S_r}^2} =  | \kappa_{Hr} | < 8\pi \, , \quad \forall \, i \rightarrow f = hh \leftrightarrow S_r S_r, \, h S_r \rightarrow h S_r \, ,\\
	&16\pi \big| a_0(i' \rightarrow f') \big|_{s\gg m_{h}^2, m_{S_r}^2} = | \lambda_{12} | < 8\pi \, , \quad \forall \, i' \rightarrow f' = S_1 S_1 \leftrightarrow S_{2} S_{2}, \, S_1 S_{2} \rightarrow S_{1} S_{2} \, , \label{eq:PU4}
\end{align}
with $r=1,2$. Additional conditions can be derived through a coupled-channel analysis, by requiring the eigenvalues of the scalar coupled-channel (symmetric) matrix $(a_0)_{2\to2}$ in the high-energy regime $s\gg m_{h}^2, m_{S_r}^2$ to be smaller (in magnitude) than $1/2$. These eigenvalues of\footnote{Each entry of the coupled-channel matrix $(a_0)_{2\to2}$ must be rescaled by a factor of $1/\sqrt{2}$ for each initial/final state of identical particles, as explained e.g. in~\cite{Logan:2022uus}.}
\begin{equation}
a_0(i \to f) \, \xrightarrow{s\gg m_{h}^2, m_{S_r}^2} \, -\frac{1}{16\pi}
\begin{pmatrix}
\frac{3 m_h^2}{ \sqrt{2} \sqrt{2} v^2} & 0 & 0 & \frac{\kappa_{H1}}{\sqrt{2} \sqrt{2}} & 0 & \frac{\kappa_{H2}}{\sqrt{2} \sqrt{2}} \\
0 & \kappa_{H1} & 0 & 0 & 0 & 0 \\
0 & 0 & \kappa_{H2} & 0 & 0 & 0 \\
\frac{\kappa_{H1}}{\sqrt{2} \sqrt{2}} & 0 & 0 & \frac{\lambda_1}{\sqrt{2} \sqrt{2}} & 0 & \frac{\lambda_{12}}{\sqrt{2} \sqrt{2}} \\
0 & 0 & 0 & 0 & \lambda_{12} & 0 \\
\frac{\kappa_{H2}}{\sqrt{2} \sqrt{2}} & 0 & 0 & \frac{\lambda_{12}}{\sqrt{2} \sqrt{2}} & 0 & \frac{\lambda_2}{\sqrt{2} \sqrt{2}}
\end{pmatrix} \, ,
\end{equation}
with $\ket{i},\ket{f} \in \{ \ket{hh}, \ket{h S_1}, \ket{h S_2}, \ket{S_1 S_1}, \ket{S_1 S_2}, \ket{S_2 S_2} \}$, are given by
\begin{align}
a_{1,2} = -\frac{\kappa_{H1,2}}{16\pi} \, , \quad a_3 = -\frac{\lambda_{12}}{16\pi} \, , \quad a_{4,5,6} = \frac{1}{32\pi v^2} \, \text{Root}\left[f(x),\, k=1,2,3\right] \, ,
\end{align}
where $\text{Root}[f(x), k]$ denotes the $k$-th root of the cubic polynomial
\begin{align}
f(x) = \, &x^3 + \left[ (\lambda_1 + \lambda_2) v^2 + 3 m_h^2 \right] x^2 + \left[ (- \kappa_{H1}^2 - \kappa_{H2}^2 + \lambda_1 \lambda_2 - \lambda_{12}^2) v^4 + 3(\lambda_1 + \lambda_2) m_h^2 v^2 \right] x \notag \\
&+ (- \kappa_{H1}^2 \lambda_2 - \kappa_{H2}^2 \lambda_1 + 2 \kappa_{H1} \kappa_{H2} \lambda_{12}) v^6 + 3 (\lambda_1 \lambda_2 - \lambda_{12}^2 ) m_h^2 v^4 \, .
\end{align}
Hence, $|a_{1, \dots, 6}|<1/2$ provide three new tree-level conditions relative to~\eqref{eq:PU1}--\eqref{eq:PU4}.

The remaining models presented in this work have a similar type of structure in the sense that there is no mixing between the visible and the dark sector -- the dark symmetries are never broken. The only experimental constraint from colliders is always the Higgs invisible width. The dark matter constraints are applied. As for the theoretical bounds we will just take the perturbativity bounds because the goal of the paper is to understand if new mass regions of the parameter space open up. Additional theoretical constraints may further limit the allowed parameter space.

\subsubsection{Parameter Space Scans and Numerical Analysis}
%
%
%
As discussed in the introduction, the extension of the SM by just one real singlet with an exact $\mathcal{Z}_2$ symmetry is highly constrained by experiment. This model has three free parameters relative to the SM, but only two of them are relevant at tree-level: the DM mass $m_{S}$ and the portal coupling $\kappa_{HS}$ (the quartic self-interaction coefficient is only relevant at loop level). Only a very heavy DM particle with a large portal coupling, or a mass close to half the Higgs mass and a portal coupling below roughly $10^{-3}$, are still allowed.
\begin{figure}[ht]
    \centering
    \includegraphics[height=0.425\textwidth]{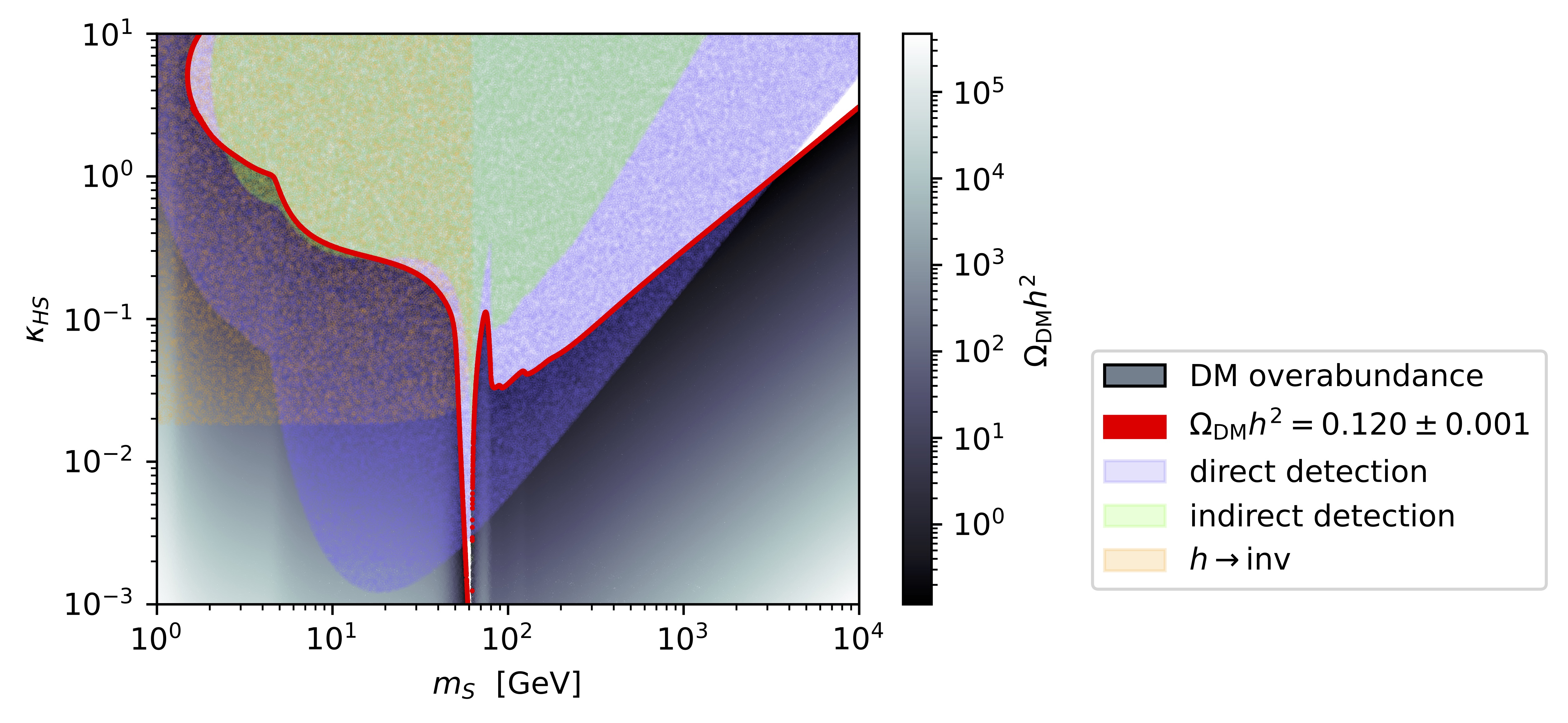}
    \caption{Experimental constraints on the real singlet extension of the SM obtained using \texttt{micrOMEGAs 6.0} for the freeze-out mechanism. The grey, purple, green and orange regions are, respectively, excluded by DM relic (over)density, direct detection, indirect detection and BR($h$$\rightarrow$inv); while the red region corresponds to the observed DM relic density.}
    \label{fig:SM+RSS scans}
\end{figure}
In Fig.~\ref{fig:SM+RSS scans} we present the allowed parameter space for the model, obtained with  \texttt{micrOMEGAs 6.0}~\cite{Alguero:2023zol} for a freeze-out DM candidate. For the invisible Higgs branching ratios, we use the LHC results from ATLAS \cite{ATLAS:2019cid} and CMS \cite{CMS:2018yfx}. For the indirect detection constraints, both \texttt{micrOMEGAs 6.0} and the MADHAT \cite{Boddy:2018qur,Boddy:2019kuw} software along with data from dwarf galaxies presented in Ref. \cite{Boddy:2019qak} was used. In all remaining plots the DM bounds have the observed DM relic density $\Omega_{\text{DM}}^{\text{obs}}h^2=0.120 \pm 0.001$ from \textsc{Planck}~\cite{Planck:2018vyg}\footnote{Throughout this work, we require the relic density to lie within the experimental interval, under the assumption that no other DM particles exist besides those predicted by our models. If instead we treated the observed relic density as an upper bound, the allowed parameter space regions would (in general) be larger.} and are in agreement with the DD bounds from XENON1T~\cite{XENON:2018voc}, DarkSide-50~\cite{DarkSide:2018bpj}, PICO-60~\cite{PICO:2019vsc}, CRESST-III~\cite{CRESST:2019jnq}, PandaX-4T~\cite{PandaX-4T:2021bab} and LUX-ZEPLIN (LZ)~\cite{LZ:2022lsv}.
Excluding the resonant region, the allowed region starts at a mass of about $3500$ GeV, clearly out of the LHC reach in any DM production channel. This value is explicitly seen in the intersection between the correct DM relic density curve and the DD bound.

\begin{figure}[h]
    \centering
    \includegraphics[width=\textwidth]{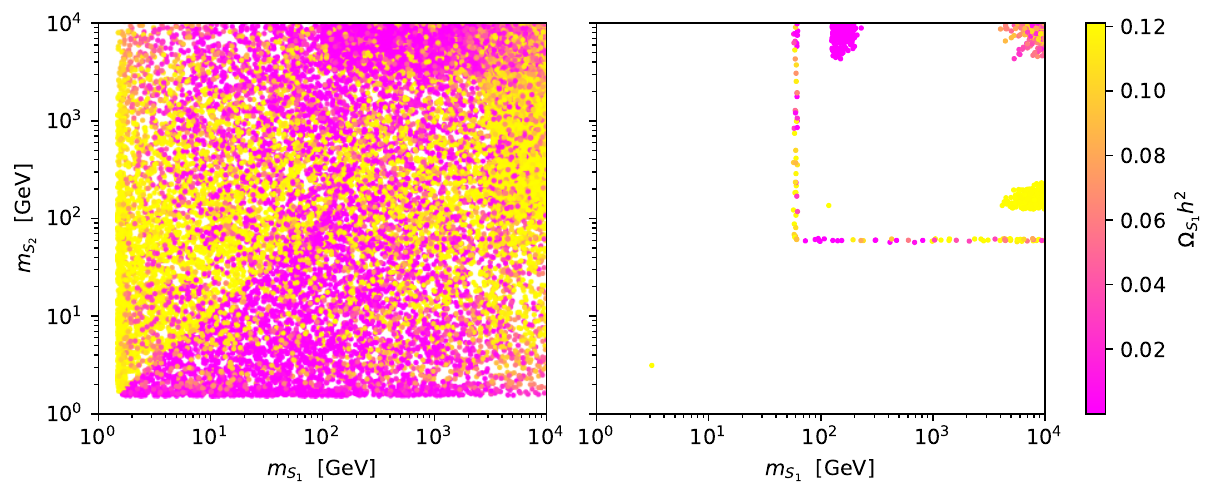}
    \caption{Allowed points for the two real singlets extension of the SM, obtained using \texttt{micrOMEGAs 6.0}. The left panel shows the parameter space points that correspond to the observed DM relic density, and the right panel
    present the points that also fulfill the DD constraints. The colour bar shows the relic density of the lighter DM particle,  $\Omega_{S_1}  h^2$.}
    \label{fig:SM+2RSS scans (relic)}
\end{figure}
Let us now move to the discussion of the two real singlets extension of the SM with two independent $\mathcal{Z}_2$ symmetries. This model has seven free parameters relative to the SM, but only five of them are relevant at tree-level: the DM masses $m_{S_1}$ and $m_{S_2}$, the portal couplings $\kappa_{H1}$ and $\kappa_{H2}$, and the inter-dark coefficient $\lambda_{12}$. The two quartic self-interaction coefficients $\lambda_1$ and $\lambda_2$ are only relevant for loop corrections in perturbation theory. We scanned the free parameter space with \texttt{micrOMEGAs 6.0}~\cite{Alguero:2023zol}, assuming both scalar DM particles were thermally produced according to the freeze-out mechanism.

Taking into account all the constraints described previously, a scan in the parameter space leads to the allowed parameter space regions presented
in Fig.~\ref{fig:SM+2RSS scans (relic)}. Without loss of generality, due to the $S_1 \leftrightarrow S_2$ symmetry of our model, we have ordered the masses choosing $S_1$ as the lighter DM particle.
On the left panel we show the points with the correct relic density corresponding to the sum of the two relic densities, that is $\Omega_{\text{DM}}h^2 = (\Omega_{S_1} + \Omega_{S_2})h^2$. Points in the right panel also passed DD constraints,
and the colour bar presents the relic density $\Omega_{S_1}  h^2$. 

There are essentially three regions of allowed parameter space. One where the two DM particles are heavy, which is the high mass region allowed in the one real singlet case. The second region
is the one where one of the DM masses is in the resonance pole of the $s$-channel Higgs propagator, $m_{S_{1,2}} \approx \sqrt{s}/2 = m_h / 2$. The resonant behaviour of the cross section allows for a smaller portal coupling of the lighter DM candidate. The small coupling is instrumental in evading the bounds from DD. With respect to the case with one singlet there is more freedom in the portal couplings values, but they still have to be of the same order of magnitude as in the singlet case. 

Finally, the third region only occurs in this extension. In this case, the allowed mass ranges are $m_{S_1} \in [124.8, 230.0]$ GeV and $m_{S_2} \in [4321.0, 9977.0]$ GeV. Hence, a new mass region with potential to be probed at the next LHC run opens up. Still, the heavier particle is again considerably suppressed in LHC production. Also, as shown in Fig.~\ref{fig:SM+2RSS scans (relic)} in the colour bar, the lighter DM particle $S_1$ has a very small fraction of the observed relic density $\Omega_{S_1} h^2 \sim [10^{-8}, 10^{-7}]$, so that $\Omega_{\text{DM}}h^2 = (\Omega_{S_1} + \Omega_{S_2})h^2 \approx \Omega_{S_2} h^2$. There is a fourth region, where $S_1$ is just interchanged with $S_2$.

\begin{figure}[h]
    \centering
    \includegraphics[width=\textwidth]{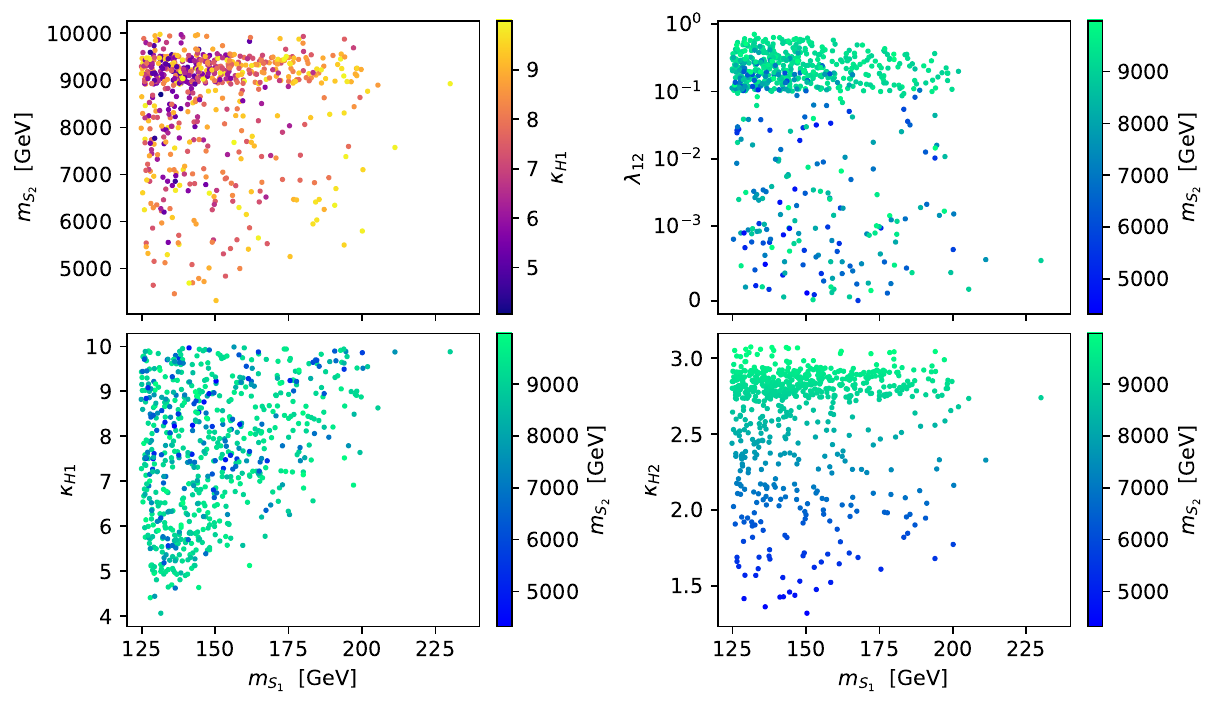}
    \caption{One-light-one-heavy $m_{S_1}<m_{S_2}$ scenario. Allowed parameter space for the two singlets extension of the SM obtained using \texttt{micrOMEGAs 6.0} for several projections, with the lightest DM particle mass on the $x$-axis. The other four relevant parameters are shown on the $y$-axis or in the colour bar. We only show the points in the new mass window where $S_1$ is the lightest DM candidate.}
    \label{fig:SM+2RSS scans}
\end{figure}

In Fig.~\ref{fig:SM+2RSS scans} we show the allowed parameter space for several projections with the mass of the lightest DM particle on the $x$-axis. The other four relevant parameters are shown either on the $y$-axis or in the colour bar. The allowed ranges for the three quartic portal coupling constants are $\kappa_{H1} \in [4.066, 9.986]$,   $\kappa_{H2} \in [1.321, 3.074]$, and $\lambda_{12} \in [0, 0.7093]$. The most important features to notice is the large values of the portal couplings to the visible sector, and in particular the one of the lightest DM particle to the Higgs field.
Besides the large portal couplings to the Higgs the two dark sectors have to be connected by a small $\lambda_{12}$.

\begin{figure}[!]
    \centering
    \includegraphics[width=\textwidth]{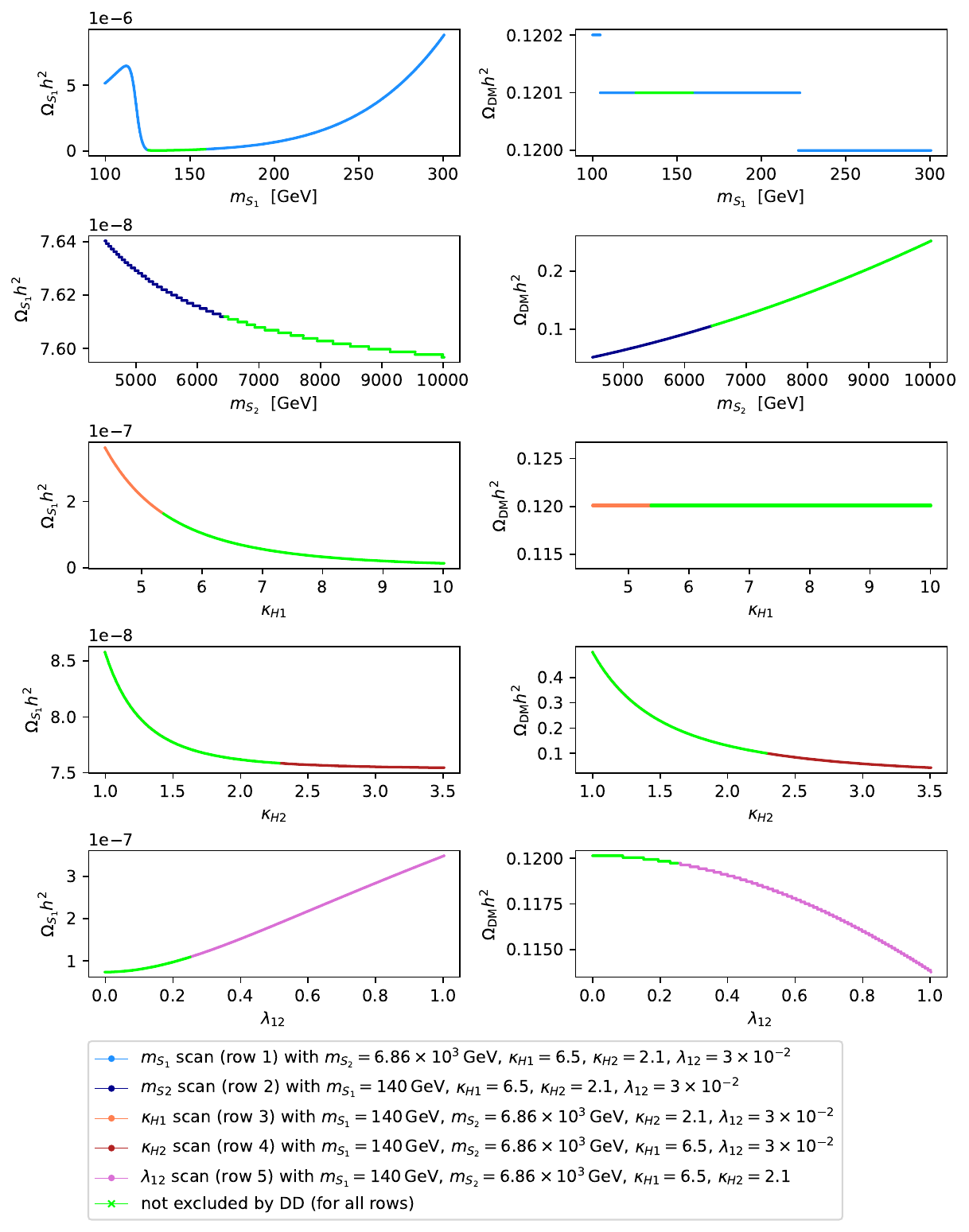}
    \caption{Analysis of the $m_{S_1}<m_{S_2}$ case with all parameters (but one) fixed. Each row scans one parameter -- $m_{S_1}$, $m_{S_2}$, $\kappa_{H1}$, $\kappa_{H2}$, and $\lambda_{12}$, in this order -- with all the remaining parameters fixed. Left: relic density $\Omega_{S_1} h^2$ of $S_1$; right: total relic density $\Omega_{\text{DM}} h^2 = (\Omega_{S_1} + \Omega_{S_2})h^2$ given by the sum of the relic densities from $S_1$ and $S_2$. Only the green points are not excluded by DD. $\lambda_{12}$ affects the total relic density, but this effect is very small compared with those of $m_{S_2}$, $\kappa_{H2}$.}
    \label{fig:interest case analysis}
\end{figure}

Before proceeding we will briefly review the expression of the DD cross section.
At leading order, the spin-independent (SI) cross section for the elastic scattering of a DM particle $S_r$ with a nucleon $N=p,n$ is given by (e.g., see~\cite{Jungman:1995df,Hisano:2015bma,DelNobile:2021wmp} for details)
\begin{equation}
    \sigma^{\text{SI}}(S_r N \rightarrow S_r N) = \frac{1}{\pi} \left( \frac{m_N}{m_N + m_{S_r}} \right)^2 |f_N^{\text{LO}}|^2 \, , \qquad r=1,2 \, ,
\end{equation}
where
\begin{equation}
    f_N^{\text{LO}} = \left[ \sum_{q=u,d,s} C_q^{\text{LO}} f_{T_q}^N + \sum_{Q=c,b,t} C_Q^{\text{LO}} \times \frac{2}{27} f_{T_G}^N \right] m_N = +\frac{\kappa_{Hr}}{2m_h^2} \left[\frac{6}{27} + \frac{21}{27} \sum_{q=u,d,s} f_{T_q}^N \right] m_N\,,
\end{equation}
is the DM-nucleon effective coupling, $C_q^{\text{LO}}=+\kappa_{Hr}/(2 m_h^2)$ are the DM-quark effective couplings at LO, $f_{T_{q}}^{N} \equiv \bra{N(\vec{\, p})} m_{q} \bar{q} q \ket{N(\vec{\, p})} / m_N$ is the fraction of nucleon mass attributed to a light quark $q=u,d,s$ contribution and $f_{T_{G}}^{N} \equiv 1 - \sum_{q=u,d,s} f_{T_q}^N$ is the fraction of nucleon mass attributed to the heavy quarks $Q=c,b,t$ and the gluon $g$.

We will now examine the reasons for the existence of the low mass region in more detail. 
In Fig.~\ref{fig:interest case analysis} we show the relic density of the lightest DM particle as a function of one of the relevant parameters at a time.  Each row scans one parameter -- $m_{S_1}$, $m_{S_2}$, $\kappa_{H1}$, $\kappa_{H2}$ and $\lambda_{12}$, in this order -- with all the remaining parameters fixed. The green lines correspond to the scenarios where DD is allowed. As a general trend we see the relic density decreasing steeply as soon as the channel $S_1 S_1 \to hh$ opens. This decrease in the relic density reaches a point where DD is also allowed. As this is an $s$-channel process the cross section decreases with increasing $m_{S_1}$, and the relic density becomes again disallowed by the Planck constraints.

It is clear from these figures that the total relic density $\Omega_{\text{DM}}h^2 = (\Omega_{S_1} + \Omega_{S_2})h^2$, although dependent on all five parameters $m_{S_1}$, $m_{S_2}$, $\kappa_{H1}$, $\kappa_{H2}$ and $\lambda_{12}$, is only significantly affected by three of them: $m_{S_2}$, $\kappa_{H2}$ and $\lambda_{12}$. This was already expected, since $\Omega_{\text{DM}}h^2 \approx \Omega_{S_2} h^2$ is determined by the relic density of the heavier DM particle $S_2$. In particular, $\Omega_{\text{DM}}h^2$ increases with $m_{S_2}$ and decreases with $\kappa_{H2}$ and $\lambda_{12}$. If we look again at the left plot in Fig.~\ref{fig:SM+2RSS scans (relic)}
we see that if it was not for the DD bounds the relic density could be equally distributed between the two DM particles. However, after DD bounds are taken into account, the heavy particle is responsible for almost all the relic density. It can be equally divided between the two if they are both heavy but if one of them is light, the latter has to have a tiny fraction of the total relic density. 

 One important point is that the inter-dark coefficient $\lambda_{12}$ which is responsible for the strength of heavy to light $S_2 S_2 \rightarrow S_1 S_1$ annihilation does not influence the total DM relic density as significantly as $m_{S_2}$ and $\kappa_{H2}$, and is therefore not constrained by it. But since it affects the $S_1$ fraction $\Omega_{S_1}/\Omega_{\text{DM}}$, it is bounded from above due to DD exclusion related to $S_1$.

We know that the heavier state $S_2$ is responsible for almost all the observed relic density. This is a possible scenario because we have already encountered it in the one singlet case. In this case, DD exclusion is completely determined by $\sigma^{\text{SI}}(S_2 N \rightarrow S_2 N)$, which must not surpass the upper bound set (under the assumption of a relic density of $0.120 \pm 0.001$) by DD experiments. This means that by increasing $m_{S_2}$ and decreasing $\kappa_{H2}$ we are promoting DD non-exclusion; however, we are also increasing $\Omega_{S_2}h^2 \approx \Omega_{\text{DM}}h^2$, thus simultaneously promoting relic over-density. 


For the lighter DM state $S_1$,  DD exclusion is not determined by $\sigma^{\text{SI}}(S_1 N \rightarrow S_1 N)$ but rather by the bound on  $\sigma^{\text{SI}}(S_1 N \rightarrow S_1 N) \times \Omega_{S_1}/\Omega_{\text{DM}}$. DD exclusion due to the lighter DM particle $S_1$ is mainly determined by $\Omega_{S_1}h^2$. Hence, evading DD exclusion related to $S_1$ requires a low fraction $\Omega_{S_1}/\Omega_{\text{DM}}$, and consequentially, low $m_{S_1}$, high $\kappa_{H1}$ and low $\lambda_{12}$ (see Fig.~\ref{fig:interest case analysis}).

\begin{figure}[!ht]  
\centering
   \includegraphics[width=\textwidth]{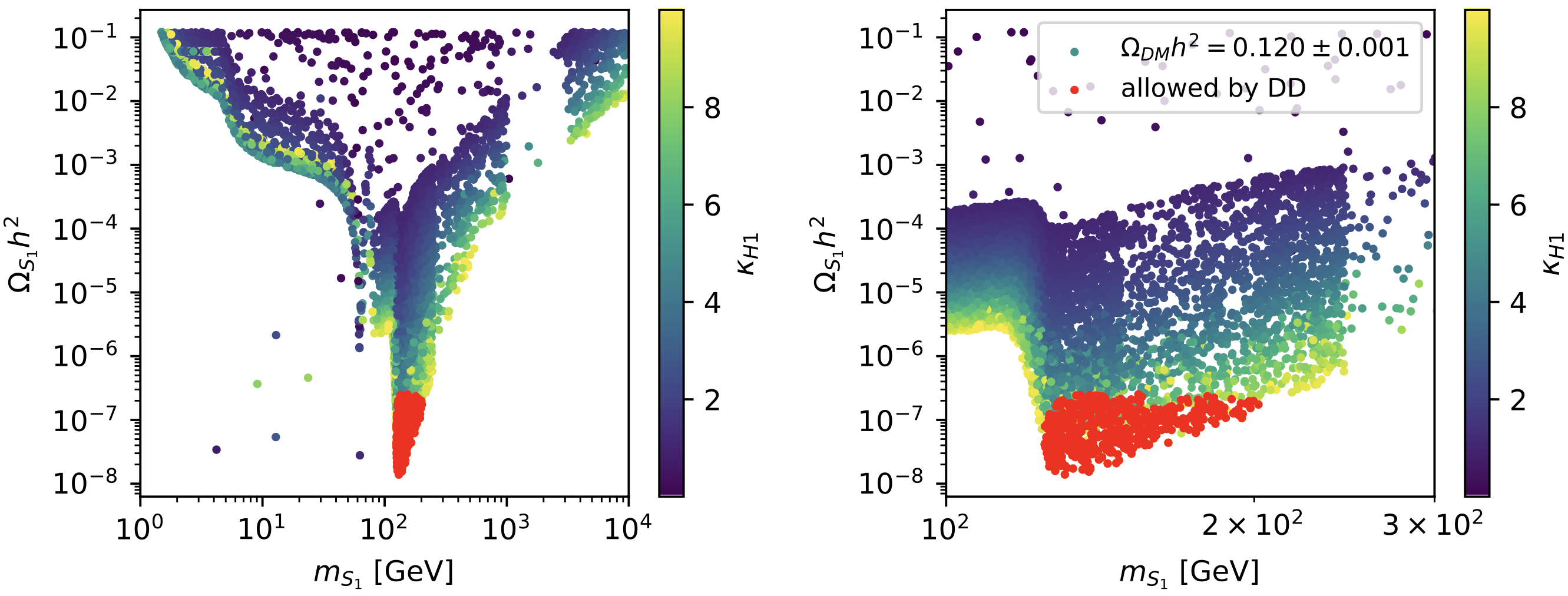} 
  \caption{Relic density of $S_1$ as a function of its mass. The colour bar shows how the coupling $\kappa_{H1}$ varies. All points have the correct DM abundance. The red points are allowed by DD experiments. The plot on the right is a zoom on the region of interest.}
  \label{relic-masses}
\end{figure}

In Fig.~\ref{relic-masses} we present the relic density of the lightest DM particle as a function of its mass.
\begin{figure}[!ht]  
\centering
   \includegraphics[width=\textwidth]{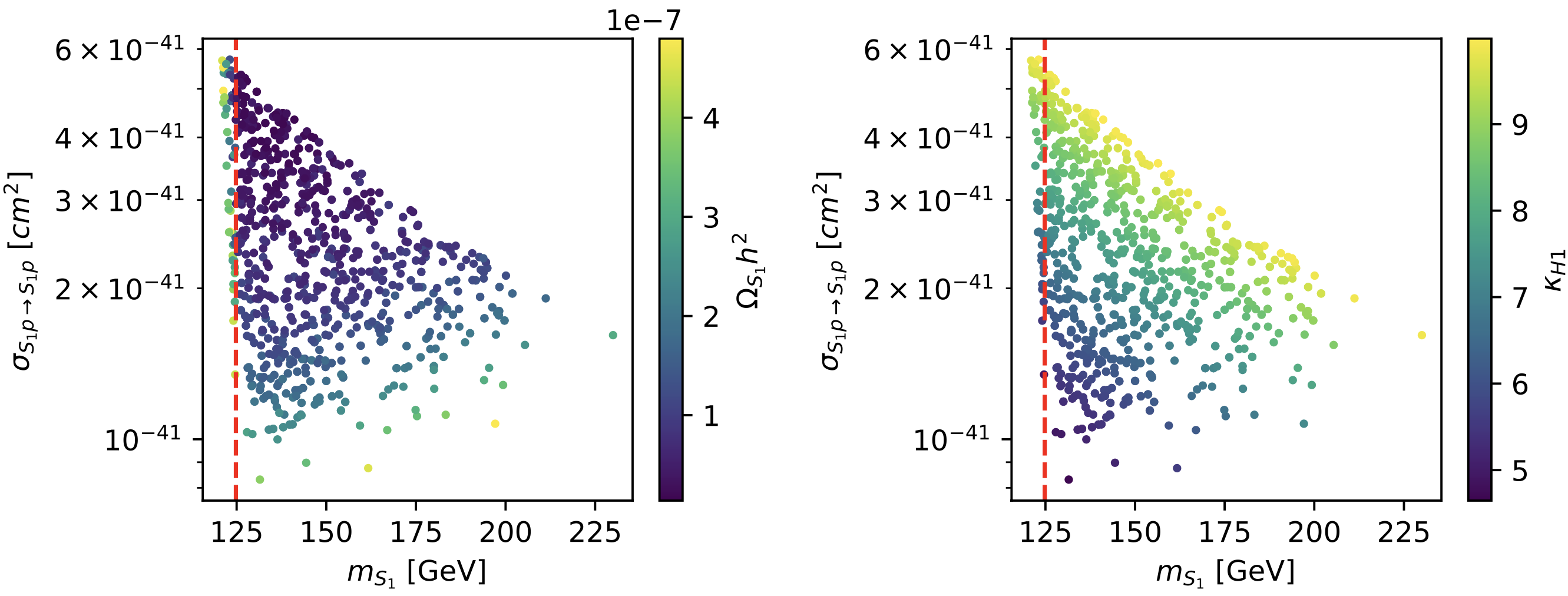} 
  \caption{Spin-independent scattering cross section of $S_1$-proton as a function of its mass. On the left panel, the colour bar represents the relic density of $S_1$. On the right panel, the colour bar represents the portal coupling $\kappa_{H1}$. All points have the correct DM abundance. The red dashed line is fixed  at the Higgs boson mass ($125$ GeV).}
    \label{cs-dd}
\end{figure}
The region of allowed points has a minimum for the relic density of $S_1$ of about $3.32 \times 10^{-7}$. The right panel zooms in the mass region of interest.
%
Regarding DD non-exclusion related to the lighter DM particle $S_1$, a low $S_1$ relic density compensates for a large $S_1$-nucleon cross section.

In Fig.~\ref{cs-dd} we present the spin-independent scattering cross section of $S_1$-proton as a 
\begin{figure}[!ht]
    \centering
    \includegraphics[width=\textwidth]{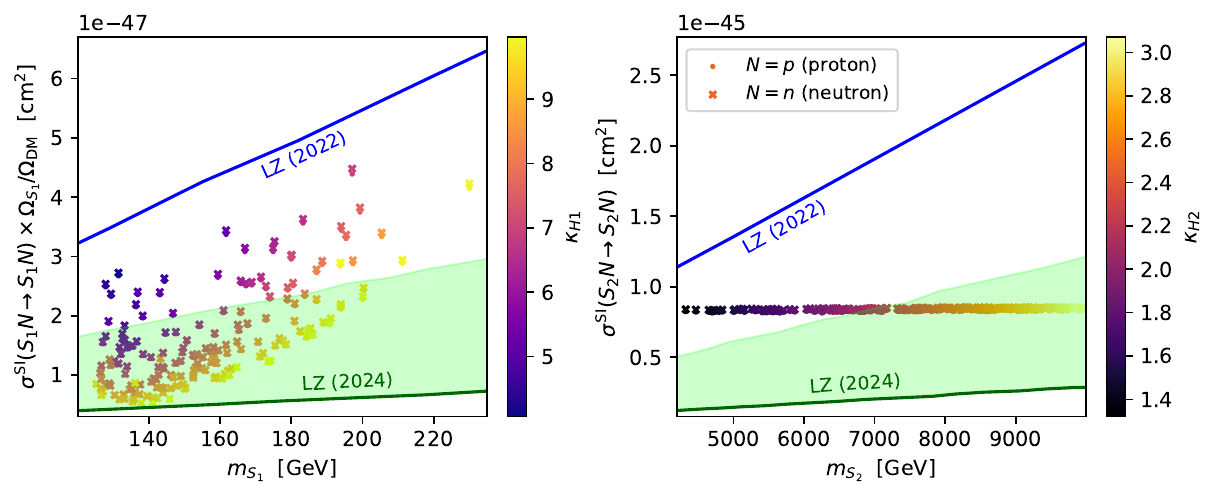}
    \caption{Spin-independent cross section of DM-nucleon elastic scattering multiplied by the corresponding fraction of DM relic density $\Omega_{S_r} / \Omega_{\text{DM}}$ ($r=1,2$), for both the lighter $S_1$ (left panel) and the heavier $S_2$ (right panel) DM particle candidates. In the right panel we do not show the fraction because it is close to one. The colour bar represents the value of the respective portal coupling.
    The blue and green (solid) lines correspond to the LUX-ZEPLIN (LZ) experimental upper limits on the WIMP-nucleon elastic scattering SI cross section from 2022 and 2024 results, respectively. The (shaded green) experimental uncertainty band from the LZ 2024~\cite{LZ:2022lsv}  results is displayed as well.}
    \label{fig:SM+2RSS DD prospects}
\end{figure}
function of the mass of $S_1$, $m_{S_1}$. On the left panel, the colour bar represents the relic density of $S_1$. On the right panel, the colour bar represents the portal coupling $\kappa_{H1}$. All points have the correct DM abundance. The allowed region starts after the red dashed line, at a mass of $124.8$ GeV. One can see that close to the red dashed line, on the left upper corner, in the region below $124.8$ GeV, there are points not allowed because the relic density of $S_1$ is still too large. 
However, after we cross the red line, the relic density decreases to the point where, even when it is multiplied by a very large scattering cross-section, the points are not excluded by the DD constraints. On the right plot we can clearly see the upper bound on the portal coupling. 

In Fig.~\ref{fig:SM+2RSS DD prospects} we show the spin-independent cross sections of DM-nucleon elastic scattering $S_r N \rightarrow S_r N$, $N=p,n$. In the left plot we show the cross section for the lighter DM candidate multiplied by its DM fraction. The right plot shows the cross section for the heavier DM particle, which has a DM fraction close to one. 
The points are shown both for proton ($p$) and for neutron ($n$) elastic scattering.   The blue and green solid lines correspond to the LUX-ZEPLIN experimental upper limits on the WIMP-nucleon elastic scattering SI cross section from 2022 and 2024 results, respectively. The shaded green experimental uncertainty band from the LZ 2024~\cite{LZ:2022lsv}  results is displayed as well. The plot shows that although the new region for this model was still allowed with the old LZ results, the points that survive are now at most in the uncertainty band of the new LZ results. This means they will most certainly be probed at the next DD experiment.


\subsection{One $\mathcal{Z}_2$ Symmetry}

Let us now discuss the two singlet extension with just one $\mathcal{Z}_2$ symmetry $S'_1 \to -S'_1$, $S'_2 \to -S'_2$, $\text{SM} \to +\text{SM}$. In this scenario, the most general renormalisable Lagrangian is given by
\begin{align}
    \notag \mathcal{L}'_{\text{SM+2RSS}} \, = \: &\mathcal{L}_{\text{SM}} + \frac{1}{2} (\partial_{\mu} S'_1) \partial^{\mu}S'_1 - \frac{1}{2}\mu_1^2 S'^2_1 + \frac{1}{2} (\partial_{\mu} S'_2) \partial^{\mu}S'_2 - \frac{1}{2} \mu_2^2 S'^2_2 - \frac{\lambda_1}{4!} S'^4_1 - \frac{\lambda_2}{4!} S'^4_2 \\
    \notag &- \frac{\kappa_{H1}}{2}S'^2_1 \Phi^{\dagger} \Phi
    - \frac{\kappa_{H2}}{2}S'^2_2 \Phi^{\dagger} \Phi - \frac{\lambda_{12}}{4} S'^2_1 S'^2_2 \\
    &- \mu_{12}^2 S'_1 S'_2 - \frac{\lambda_3}{3!} S'^3_1 S'_2 - \frac{\lambda_4}{3!} S'_1 S'^3_2 -\kappa_{H12} S'_1 S'_2   \Phi^{\dagger} \Phi \, , 
    \label{eq:SM+2RSSprime}
\end{align}
where the last four terms are new relative to the previous Lagrangian~\eqref{eq:SM+2RSS} with two independent $\mathcal{Z}_2$ symmetries. Again, we define $\Phi = \begin{pmatrix} G^+, & ( v + h + i G^0)/\sqrt{2} \end{pmatrix}^{\text{T}}$ as the SM Higgs doublet. 
The new terms induce mixing between the DM states. In fact, the $S'_r$ fields are now gauge eigenstates, and the corresponding mass eigenstates $\chi_r$ are given by
\begin{equation}
    \begin{pmatrix}
        \chi_1 \\ \chi_2
    \end{pmatrix} = U^{\text{T}}(\alpha) \begin{pmatrix}
        S'_1 \\ S'_2
    \end{pmatrix} \, , \quad \text{where} \quad
    \begin{cases}
        U(\alpha) = 
    \begin{pmatrix}
        \cos{\alpha} & +\sin{\alpha} \\ -\sin{\alpha} & \cos{\alpha}
    \end{pmatrix} \in \text{SO}(2) \\
        \tan{(2\alpha)} = \frac{2 \big( \mu_{12}^2 + \frac{\kappa_{H12}v^2}{2} \big) }{(\mu_2^2 - \mu_1^2) + \frac{(\kappa_{H2}-\kappa_{H1})v^2}{2}}
    \end{cases} .
\end{equation}
In the mass basis, the Lagrangian can be written in the unitary gauge as
\begin{align}
    \notag \mathcal{L}'_{\text{SM+2RSS}} \, = \: &\mathcal{L}_{\text{SM}} + \sum_{r=1}^{2} \left[ \frac{1}{2}(\partial_\mu \chi_r)\partial^\mu \chi_r - \frac{1}{2}m_{\chi_r}^2 \chi_r^2 - \frac{\overline{\lambda}_r}{4!} \chi_r^4 \right] - \frac{\overline{\lambda}_{12}}{4}\chi_1^2 \chi_2^2 - \frac{\overline{\lambda}_{3}}{3!}\chi_1^3 \chi_2 - \frac{\overline{\lambda}_{4}}{3!}\chi_1 \chi_2^3 \\
    &-\sum_{r=1}^2 \left[ \frac{\overline{\kappa}_{Hr} v}{2} h \chi_r^2 +\frac{\overline{\kappa}_{Hr}}{4} h^2 \chi_r^2 \right] - \overline{\kappa}_{H12} v h \chi_1 \chi_2 - \frac{\overline{\kappa}_{H12}}{2} h^2 \chi_1 \chi_2 \, ,
\end{align}
where the masses of the physical scalar particles $\chi_r$ ($r=1,2$) are given by (with $m_{\chi_1} <  m_{\chi_2}$)
\begin{equation}
    m_{\chi_{1,2}}^2 = \frac{\sum_{r=1}^2 \big( \mu_r^2 + \frac{\kappa_{Hr} v^2}{2} \big) \mp \sqrt{\big( \mu_2^2 - \mu_1^2 + \frac{(\kappa_{H2}-\kappa_{H1})v^2}{2} \big)^2 +4 \big( \mu_{12}^2 + \frac{\kappa_{H12}v^2}{2} \big)^2 }}{2} \, ,
\end{equation}
and the redefined couplings of the physical fields are
\begin{align}
    \overline{\lambda}_{1} = \, &\lambda_1 c^4_\alpha + \lambda_2 s^4_\alpha + 6 \lambda_{12} s^2_\alpha c^2_\alpha -4(\lambda_3 c^2_\alpha + \lambda_4 s^2_\alpha)s_\alpha c_\alpha\\
    \overline{\lambda}_{2} = \, &\lambda_1 s^4_\alpha + \lambda_2 c^4_\alpha + 6 \lambda_{12} s^2_\alpha c^2_\alpha +4(\lambda_3 s^2_\alpha + \lambda_4 c^2_\alpha)s_\alpha c_\alpha\\
    \overline{\lambda}_{12} = \, &(\lambda_1 + \lambda_2)s^2_\alpha c^2_\alpha + \lambda_{12} (1-6 s^2_\alpha c^2_\alpha) + 2(\lambda_3 - \lambda_4)s_\alpha c_\alpha(1-2s^2_\alpha) \\
    \overline{\lambda}_{3} = \, &(\lambda_1c^2_\alpha -\lambda_2 s^2_\alpha) s_\alpha c_\alpha +3\lambda_{12}s_\alpha c_\alpha (1-2c^2_\alpha) + \lambda_3 c^2_\alpha (1-4s^2_\alpha) + \lambda_4 s^2_\alpha (4c^2_\alpha -1)\\
    \overline{\lambda}_{4} = \, &(\lambda_1s^2_\alpha -\lambda_2 c^2_\alpha) s_\alpha c_\alpha -3\lambda_{12}s_\alpha c_\alpha (1-2c^2_\alpha) + \lambda_3 s^2_\alpha (4c^2_\alpha -1) + \lambda_4 c^2_\alpha (1-4 s^2_\alpha)\\
    \overline{\kappa}_{H1} = \, &\kappa_{H1}c^2_\alpha + \kappa_{H2}s^2_\alpha - 2\kappa_{H12}s_\alpha c_\alpha\\
    \overline{\kappa}_{H2} = \, &\kappa_{H1}s^2_\alpha + \kappa_{H2}c^2_\alpha + 2\kappa_{H12}s_\alpha c_\alpha\\
    \overline{\kappa}_{H12} = \, &(\kappa_{H1} - \kappa_{H2})s_\alpha c_\alpha + \kappa_{H12}(1-2s^2_\alpha)\, ,
\end{align}
with $s_\alpha \equiv \sin{\alpha}$, $c_\alpha \equiv \cos{\alpha}$. Since $\chi_1$ is the lightest particle from the dark sector, it will be the DM candidate. This model has eleven free parameters relative to the SM, which can be chosen as: $m_{\chi_1}$, $m_{\chi_2}$, $\overline{\kappa}_{H1}$, $\overline{\kappa}_{H12}$, $\overline{\kappa}_{H2}$, $\overline{\lambda}_{12}$, $\overline{\lambda}_1$, $\overline{\lambda}_2$, $\overline{\lambda}_3$, $\overline{\lambda}_4$ and $\alpha$.

\begin{figure}[!ht]
    \centering
    \includegraphics[width=\textwidth]{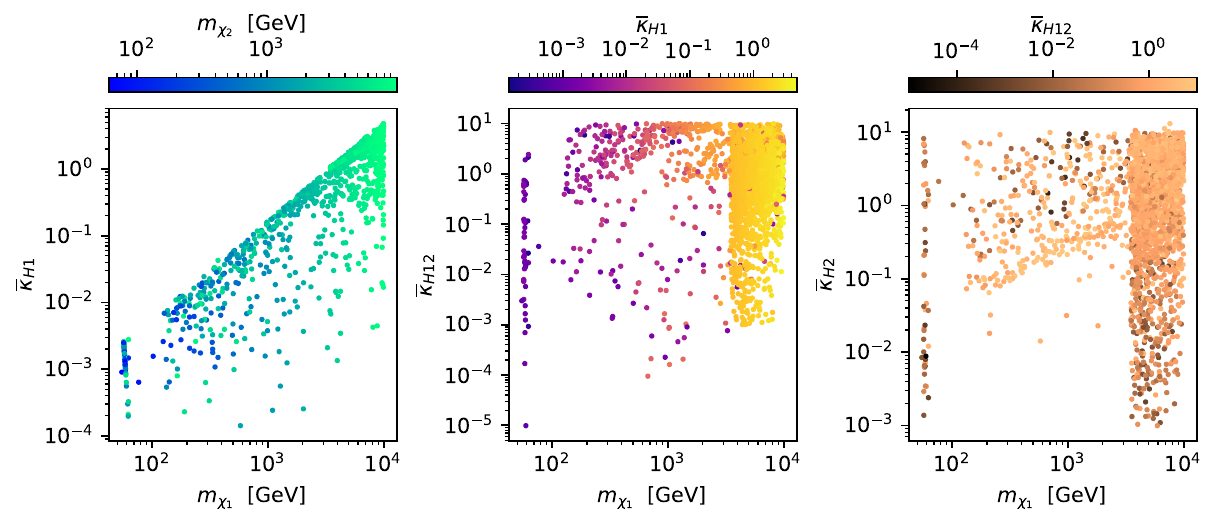}
    \caption{ $\overline{\kappa}_{H1}$,  $\overline{\kappa}_{H2}$ and  $\overline{\kappa}_{H12}$ as a function of the DM mass with three other variables in the colour bar. The points have passed all relevant bounds.}
    \label{fig:aaa}
\end{figure}

\begin{figure}[!ht]
    \centering
    \includegraphics[width=\textwidth]{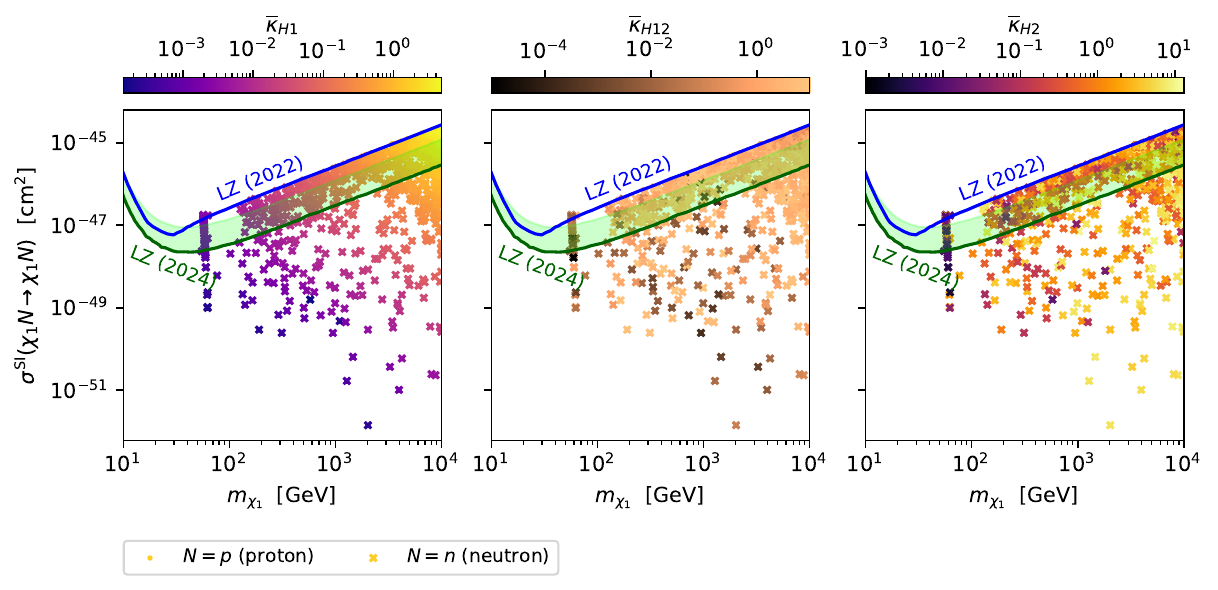}
    \caption{DD cross sections as a function of the DM mass with  $\overline{\kappa}_{H1}$,  $\overline{\kappa}_{H2}$ and  $\overline{\kappa}_{H12}$ in the colour bar. Points have survived all constraints and have the correct relic density within the experimental error. 
    The LZ bounds from 2022 and 2024 are shown as well.}
    \label{fig:bbb}
\end{figure}

What is new in this model with just one dark symmetry is that while there is only one coupling responsible for DD, $\overline{\kappa}_{H1}$, the relic density has contributions from all portal couplings, that is, $\overline{\kappa}_{H1}$, $\overline{\kappa}_{H2}$ and $\overline{\kappa}_{H12}$. Therefore, contrary to the two symmetries scenario, we can have a small $\overline{\kappa}_{H1}$, enforcing the DD bound, and make the two other relevant couplings $\overline{\kappa}_{H2}$ and $\overline{\kappa}_{H12}$ large enough so that the relic density is in agreement with the experimental value.

This behaviour is shown in Fig.~\ref{fig:aaa} where we plot the couplings $\overline{\kappa}_{H1}$,  $\overline{\kappa}_{H2}$ and  $\overline{\kappa}_{H12}$ as a function of the DM mass. 
It is clear that in order to have a small $\overline{\kappa}_{H1}$ we need to have a sufficiently large $\overline{\kappa}_{H2}$ or $\overline{\kappa}_{H12}$ (or both).
Also, $\overline{\kappa}_{H1}$ has to be smaller for lighter DM and grows with the DM mass in agreement with the DD bound. In the other plots we note that the other two portal couplings can be large, meaning they are not affected by that bound. 

In Fig.~\ref{fig:bbb} we present the DD cross sections as a function of the DM mass $m_{\chi_1}$ with $\overline{\kappa}_{H1}$,  $\overline{\kappa}_{H2}$ and  $\overline{\kappa}_{H12}$ in the colour bar. As expected, in the low mass region $\overline{\kappa}_{H1}$
has to be small while at least one of the corresponding values of  $\overline{\kappa}_{H2}$ and $\overline{\kappa}_{H12}$ has to be large as previously discussed.


\section{Three Real Singlets Extension of the SM}
\label{sec:3}

In this section we go back to the case of independent symmetries, now with an extension of the SM with three real scalar singlet fields $S_1, S_2, S_3 \sim (\mathbf{1}, \mathbf{1}, 0)$, with a Lagrangian invariant under $\mathcal{Z}^{(1)}_2 \times \mathcal{Z}^{(2)}_2 \times \mathcal{Z}^{(3)}_2$ transformations,
\begin{align}
    \mathcal{Z}^{(1)}_2 : \quad &S_1 \rightarrow - S_1, \quad S_2 \rightarrow + S_2, \quad S_3 \rightarrow + S_3, \quad \text{SM} \rightarrow +\text{SM}\\
    \mathcal{Z}^{(2)}_2 : \quad &S_1 \rightarrow + S_1, \quad S_2 \rightarrow - S_2, \quad S_3 \rightarrow + S_3, \quad \text{SM} \rightarrow +\text{SM}\\
    \mathcal{Z}^{(3)}_2 : \quad &S_1 \rightarrow + S_1, \quad S_2 \rightarrow + S_2, \quad S_3 \rightarrow - S_3, \quad \text{SM} \rightarrow +\text{SM}\, .
\end{align}
The most general renormalisable and $\text{SU}(3)_c \times \text{SU}(2)_L \times \text{U}(1)_Y \times \mathcal{Z}_2^{(1)} \times \mathcal{Z}_2^{(2)} \times \mathcal{Z}_2^{(3)}$ invariant Lagrangian is given by
\begin{align}
    \notag \mathcal{L}_{\text{SM+3RSS}} \, = \: &\mathcal{L}_{\text{SM}} + \sum_{r=1}^{3} \bigg[ \frac{1}{2} (\partial_{\mu} S_r) \partial^{\mu}S_r - \frac{1}{2}\mu_r^2 S_r^2 - \frac{\lambda_r}{4!} S_r^4 - \frac{\kappa_{Hr}}{2}S_r^2 \Phi^{\dagger} \Phi  \bigg]\\
    &- \frac{\lambda_{12}}{4} S_1^2 S_2^2
    - \frac{\lambda_{23}}{4} S_2^2 S_3^2
    - \frac{\lambda_{31}}{4} S_3^2 S_1^2 \, . \label{eq:SM+3RSS}
\end{align}
From the sixteen $\text{SU}(3)_c \times \text{SU}(2)_L \times \text{U}(1)_Y \times \mathcal{Z}^{(1)}_2 \times \mathcal{Z}^{(2)}_2 \times \mathcal{Z}^{(3)}_2$ invariant minimum solution sets, we consider
\begin{equation}
\braket{\Phi^\dagger \Phi}_0 = -\frac{\mu_H^2}{2\lambda_H} \equiv \frac{v^2}{2} \, , \quad \braket{S_r}_0 = 0 \, , \quad r=1,2,3  \label{eq:SM+3RSS vacuum 2}
\end{equation}
as the vacuum configuration of the model. For this minimum solution set, the three singlets do not acquire a VEV, so that the $\mathcal{Z}^{(1)}_2 \times \mathcal{Z}^{(2)}_2 \times \mathcal{Z}^{(3)}_2$ symmetry is not spontaneously broken and there is no mixing with the Higgs. Using the minimum condition $v^2=-\mu_H^2/\lambda_H $ in~\eqref{eq:SM+3RSS vacuum 2}, the scalar potential can be written in the unitary gauge as
\begin{align}
    \notag V(|\Phi|, S_{r=1,2,3}) \, = \, &\frac{1}{2}\overbrace{(2 \lambda_H v^2)}^{= \, m_h^2} h^2 + \lambda_H v h^3 + \frac{\lambda_H}{4}h^4 + \frac{\lambda_{12}}{4}S_1^2 S_2^2 + \frac{\lambda_{23}}{4}S_2^2 S_3^2 + \frac{\lambda_{31}}{4}S_3^2 S_1^2\\ &+ \sum_{r=1}^{3} \bigg[ \frac{1}{2}\underbrace{(\mu_r^2 + \frac{\kappa_{Hr}v^2}{2})}_{= \, m^2_{S_r}} S_r^2 + \frac{\lambda_{r}}{4!}S_r^4
    + \frac{\kappa_{Hr}v}{2} h S_r^2 + \frac{\kappa_{Hr}}{4} h^2 S_r^2 \bigg] \, .
\end{align}
The unbroken $\mathcal{Z}^{(1)}_2 \times \mathcal{Z}^{(2)}_2 \times \mathcal{Z}^{(3)}_2$ symmetry ensures that the $S_r$ ($r=1,2,3$) do not decay, thus being DM candidates.

\subsection{Parameter Space Scans and Numerical Analysis}
\label{sec:SM+3RSS numerical analysis}

This model has twelve free parameters relative to the SM (five more than the two singlets extension), but only nine of them are relevant at tree-level: the DM masses $m_{S_1}$, $m_{S_2}$ and $m_{S_3}$, the portal couplings $\kappa_{H1}$, $\kappa_{H2}$ and $\kappa_{H3}$, and the inter-dark coefficients $\lambda_{12}$, $\lambda_{23}$ and $\lambda_{31}$. The three quartic self-interaction coefficients $\lambda_1$, $\lambda_2$ and $\lambda_3$ are only relevant at the loop level in perturbation theory. We scanned the free parameter space with \texttt{micrOMEGAs 6.1}~\cite{Alguero:2023zol}, assuming all scalar DM particles were thermally produced via freeze-out mechanism.

\begin{figure}[!ht]
    \centering
    \includegraphics[width=0.992\textwidth]{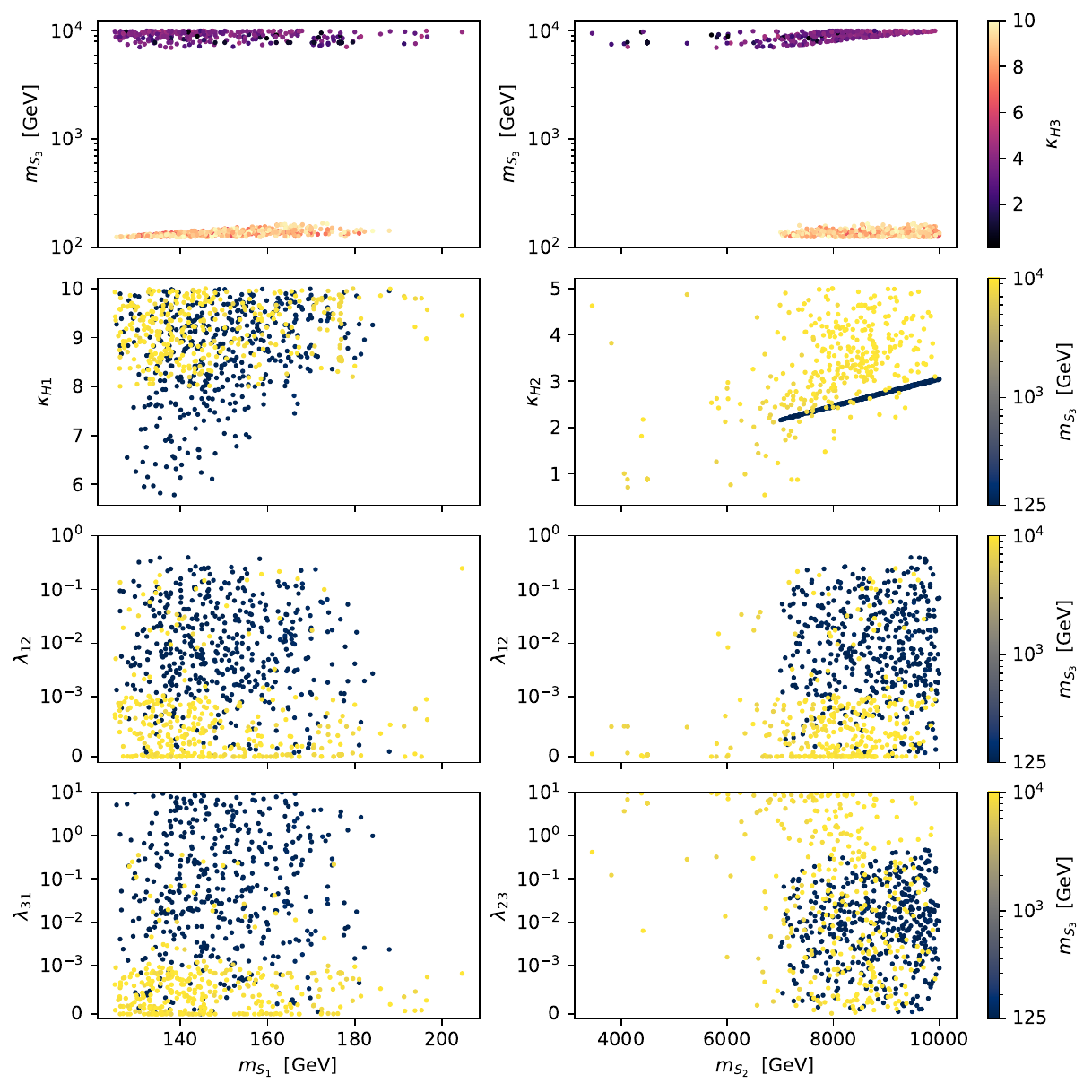}
    \caption{Experimental constraints on the SM+3RSS model~\eqref{eq:SM+3RSS}, obtained by scanning the free parameter space with \texttt{micrOMEGAs 6.1} for the freeze-out mechanism. Both panels show free parameter space points that correspond to the observed DM relic density and are not excluded by DD. Columns share the $x$-axis, and rows share the colour bar.}
    \label{fig:SM+3RSS scans}
\end{figure}

\begin{figure}[!ht]
    \centering
    \includegraphics[width=\textwidth]{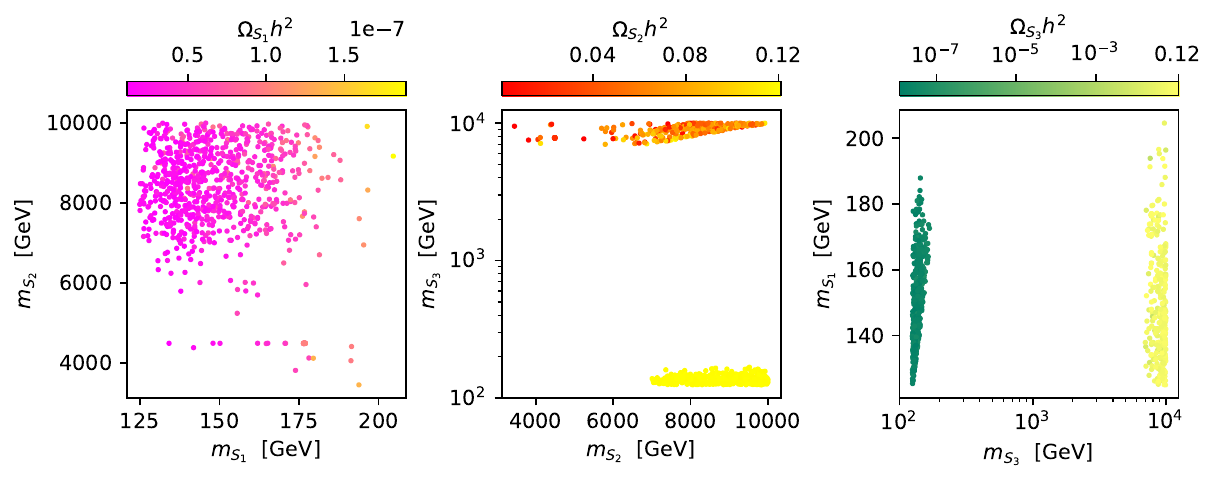}
    \caption{Experimental constraints on the SM+3RSS model~\eqref{eq:SM+3RSS}, obtained by scanning the free parameter space with \texttt{micrOMEGAs 6.1} for the freeze-out mechanism. All panels show free parameter space points that correspond to the observed DM relic density and are not excluded by DD. Each colour bar displays the relic density of one of the DM particles $S_r$ ($r=1,2,3$).}
    \label{fig:SM+3RSS scans (relic)}
\end{figure}

In the two singlets model, the full domain of parameters was scanned exhaustively and uniformly, which in principle allowed us to identify the entire allowed parameter space. However, the same was not possible for the three scalar extension due to the large number of free parameters. Instead, we used the information from the two singlet model and scanned over two targeted regions of the parameter space:
\begin{itemize}
	\item the two light (masses below 1 TeV) and one heavy (mass above 1 TeV) case region, where $S_3$ is the additional light DM particle;
	\item the one light (mass below 1 TeV) and two heavy (masses above 1 TeV) case region, where $S_3$ is the additional heavy DM particle.
\end{itemize}
This is justified by the fact that there are no new DM (co-)annihilation channels opening up in this model and therefore there are no new processes that could make DM depletion more effective. Although apparently a new channel is open, because the total number of DM particles is fixed, there is no new channel that allows for DM depletion. We imposed $m_{S_3}<m_{S_1}<m_{S_2}$ for the two-light-one-heavy case and $m_{S_1}<m_{S_2}<m_{S_3}$ for the one-light-two-heavy case regions. The coupling constants can take the usual values, but we focused on parameter configurations that could potentially be allowed by the observed relic density and DD experiments (inspired by our previous numerical analysis for the two singlet case).

The allowed parameter space of the model is shown in Figs.~\ref{fig:SM+3RSS scans} and~\ref{fig:SM+3RSS scans (relic)}. 
Fig.~\ref{fig:SM+3RSS scans} presents several projections with $m_{S_1}$ and $m_{S_2}$ on the $x$-axis of the left and right panels, respectively, and the seven remaining free parameters $m_{S_3}$, $\kappa_{H1}$, $\kappa_{H2}$, $\kappa_{H3}$, $\lambda_{12}$, $\lambda_{23}$ and $\lambda_{31}$ on the $y$-axis and colour bars. Fig.~\ref{fig:SM+3RSS scans (relic)} presents the projections in the $(m_{S_1}, m_{S_2})$, $(m_{S_2}, m_{S_3})$ and $(m_{S_3}, m_{S_1})$ planes, with the colour bars displaying the relic density of $S_1$, $S_2$ and $S_3$, respectively. Let us now proceed to the analysis of the results. 
\begin{figure}[h!]
    \centering
    \includegraphics[width=0.9\textwidth]{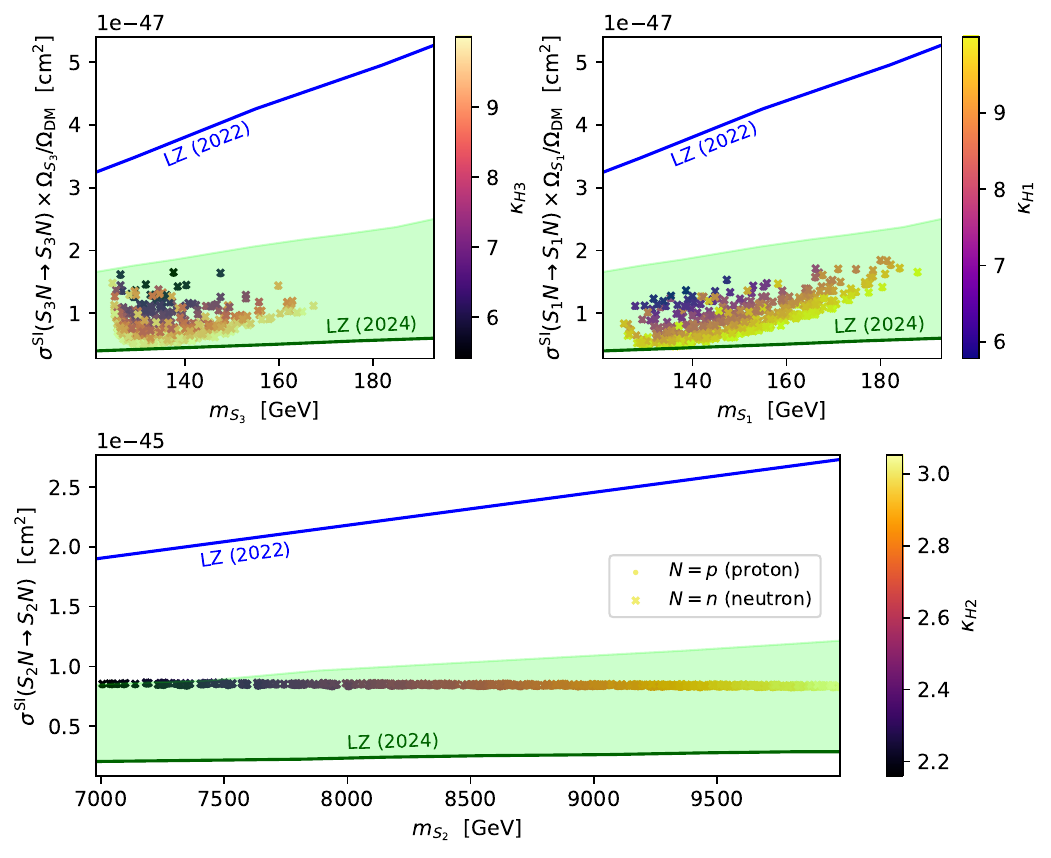}
    \caption{Two-light-one-heavy $m_{S_3}<m_{S_1}<m_{S_2}$ scenario. Spin-independent cross section of DM-nucleon elastic scattering $S_r N \rightarrow S_r N$ ($N=p,n$) multiplied by the corresponding fraction of DM relic density $\Omega_{S_r} / \Omega_{\text{DM}}$ ($r=1,2,3$), for both proton ($p$) and neutron ($n$) elastic scattering. Note that once more the DM fraction for $S_2$ is of order one. The blue and green (solid) lines correspond to the LUX-ZEPLIN (LZ) experimental upper limits on the WIMP-nucleon elastic scattering spin-independent cross section from the 2022 and 2024 results, respectively. The (shaded green) experimental uncertainty band from the LZ 2024 results is displayed as well.}
    \label{fig:SM+3RSS 2L1H DD prospects}
\end{figure}
\begin{figure}[!ht]
    \centering
    \includegraphics[width=0.9\textwidth]{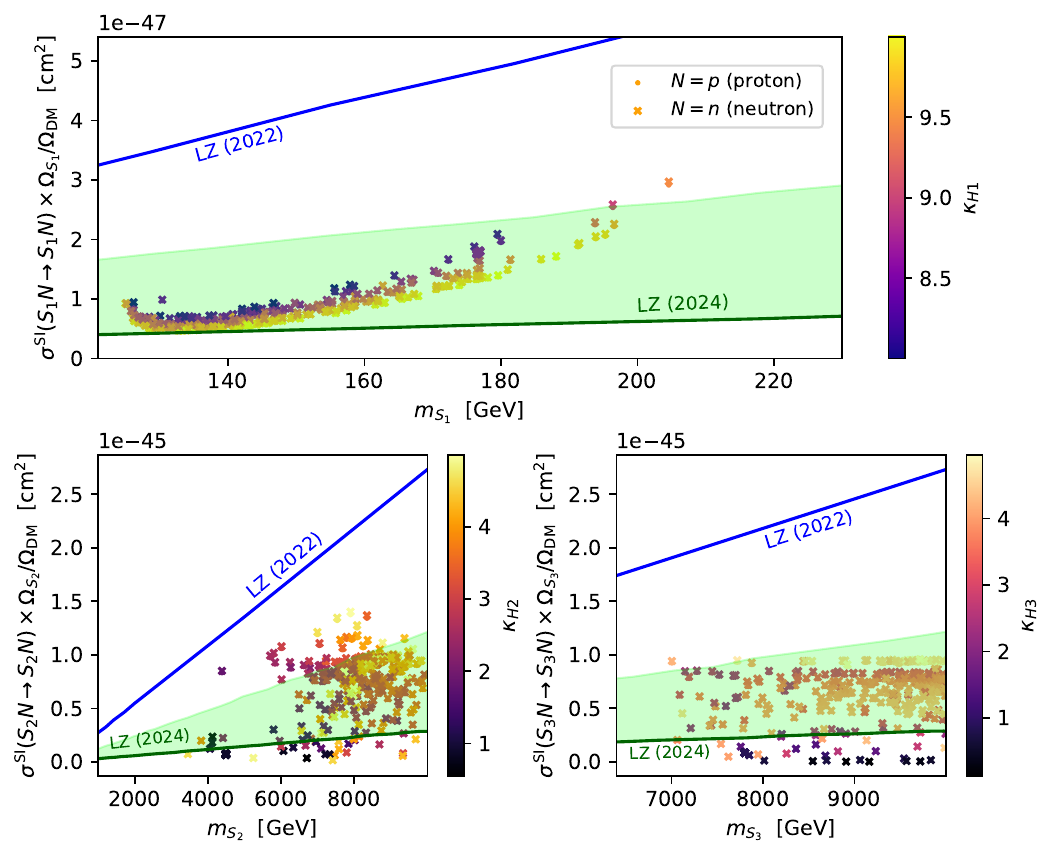}
    \caption{One-light-two-heavy $m_{S_1}<m_{S_2}<m_{S_3}$ scenario. Spin-independent cross section of DM-nucleon elastic scattering $S_r N \rightarrow S_r N$ ($N=p,n$) multiplied by the corresponding fraction of the DM relic density $\Omega_{S_r} / \Omega_{\text{DM}}$ ($r=1,2,3$) for both proton ($p$) and neutron ($n$) elastic scattering. The blue and green (solid) lines correspond to the LUX-ZEPLIN (LZ) experimental upper limits on the WIMP-nucleon elastic scattering spin-independent cross section from the 2022 and 2024 results, respectively. The (shaded green) experimental uncertainty band from the LZ 2024 results is displayed as well.}
    \label{fig:SM+3RSS 1L2H DD prospects}
\end{figure}
\begin{itemize}
    \item The \textbf{two-light-one-heavy} $\mathbf{m_{S_3} < m_{S_1} < m_{S_2}}$ \textbf{case} allowed region closely resembles the two singlet scenario. The lighter DM particles $S_1$ and $S_3$ have negligible relic densities, $\Omega_{S_3}h^2 \sim \Omega_{S_1}h^2  \sim 10^{-7}$, as shown in Fig.~\ref{fig:SM+3RSS scans (relic)}, and the heavier DM particle $S_2$ is responsible for all the observed DM relic density, i.e. $\Omega_{\text{DM}}h^2= \sum_{r=1}^3 \Omega_{S_r}h^2 \approx \Omega_{S_2} h^2$. 
    \item The \textbf{one-light-two-heavy} $\mathbf{m_{S_1} < m_{S_2} < m_{S_3}}$ \textbf{case} allowed region, in contrast, is different from the two singlets case.  As expected, the relic density of the lighter DM particle $S_1$ is negligible ($\Omega_{S_1}h^2 \sim 10^{-7}$), as shown in Fig.~\ref{fig:SM+3RSS scans (relic)}, but the two heavier DM particles $S_2$ and $S_3$ are simultaneously responsible for the observed DM relic density, i.e.
\begin{equation}
    (0.120 \pm 0.001=) \ \Omega_{\text{DM}}h^2= \sum_{r=1}^3 \Omega_{S_r}h^2 \approx (\Omega_{S_2} + \Omega_{S_3}) h^2 \, , \quad \text{with} \quad \Omega_{S_2} h^2 \sim \Omega_{S_3} h^2 \, .
\end{equation}
In addition, the DD exclusion related to the heavy DM particles $S_{2,3}$ is now determined not only by the SI cross sections, but also by the DM relic density fractions in $\sigma^{\text{SI}}(S_{2,3} N \rightarrow S_{2,3} N) \times \Omega_{S_{2,3}}/\Omega_{\text{DM}}$. Consequently, the tension between relic density and DD exclusion is alleviated for the heavy DM particles. This in turn translates into the weakening of constraints on the heavier DM particle masses $m_{S_{2,3}}$ and on the portal coefficients $\kappa_{H2,3}$ allowed values.
\end{itemize}
It should be noted that although our scans were performed for particular regions of the parameter space there are no obvious physical reasons to expect the new  allowed regions would arise. Let us now discuss what is the status of the model by looking at the latest results from DD from LZ 2024~\cite{haselschwardt2024lz}.
\begin{itemize}
    \item The \textbf{two-light-one-heavy} $\mathbf{m_{S_3} < m_{S_1} < m_{S_2}}$ \textbf{case} results are shown in Fig.~\ref{fig:SM+3RSS 2L1H DD prospects}, for the lighter $S_3, S_1$ (top panels) and heavier $S_2$ (bottom panel) DM particles. One should note that, just like in the (one-light-one-heavy) $m_{S_1}<m_{S_2}$ case of the SM+2RSS model, the heavier DM particle $S_2$ is responsible for virtually all DM relic density. Therefore it is not surprising that the results resemble very much the ones from the two singlets model with most points still inside the LZ 2024 experimental uncertainty band.
    \item The \textbf{one-light-two-heavy} $\mathbf{m_{S_1} < m_{S_2} < m_{S_3}}$ \textbf{case} results are shown in Fig.~\ref{fig:SM+3RSS 1L2H DD prospects}, for the lighter $S_1$ (top panel) and heavier $S_2,S_3$ (bottom panels) DM particles. 
    Similarly to the other scenarios the lighter DM candidate has a negligible relic density $\Omega_{S_{1}} h^2 \sim 10^{-7}$, and for masses of $m_{S_{1}}\sim[135,140]$ GeV and a high portal coefficient $\kappa_{H1}$ ($\gtrsim 9$) the points are withing the 1$\sigma$ uncertainty bound the LZ 2024 result. 
    The difference in this scenario is that the other two heavy DM states share most of the relic density. The DD cross sections for the heavy states are therefore $\sigma^{\text{SI}}(S_{2,3} N \rightarrow S_{2,3} N) \times \Omega_{S_{2,3}} / \Omega_{\text{DM}}$ .     
\end{itemize}

In conclusion, the three-real-scalar-singlet SM extension~\eqref{eq:SM+3RSS} provides an allowed free parameter space region -- the (one-light-two-heavy) $m_{S_1}<m_{S_2}<m_{S_3}$ case allowed region -- where for some of its points, the $\sigma^{\text{SI}}(S_r N \rightarrow S_r N) \times \Omega_{S_r} / \Omega_{\text{DM}}$ model predictions are simultaneously within the 1-$\sigma$ uncertainty limit of LZ 2024 and therefore not excluded by DD.

\section{LHC Dark Matter Searches}
\label{sec:4} 

In all the models presented previously, there is the possibility of having DM masses of the order of 100 GeV with a large portal coupling. But the question of detecting scalar DM from a typical portal coupling is a much more general
question that we will answer in this section. 


\begin{figure}[!ht]
    \centering
    \includegraphics[width=0.95\textwidth]{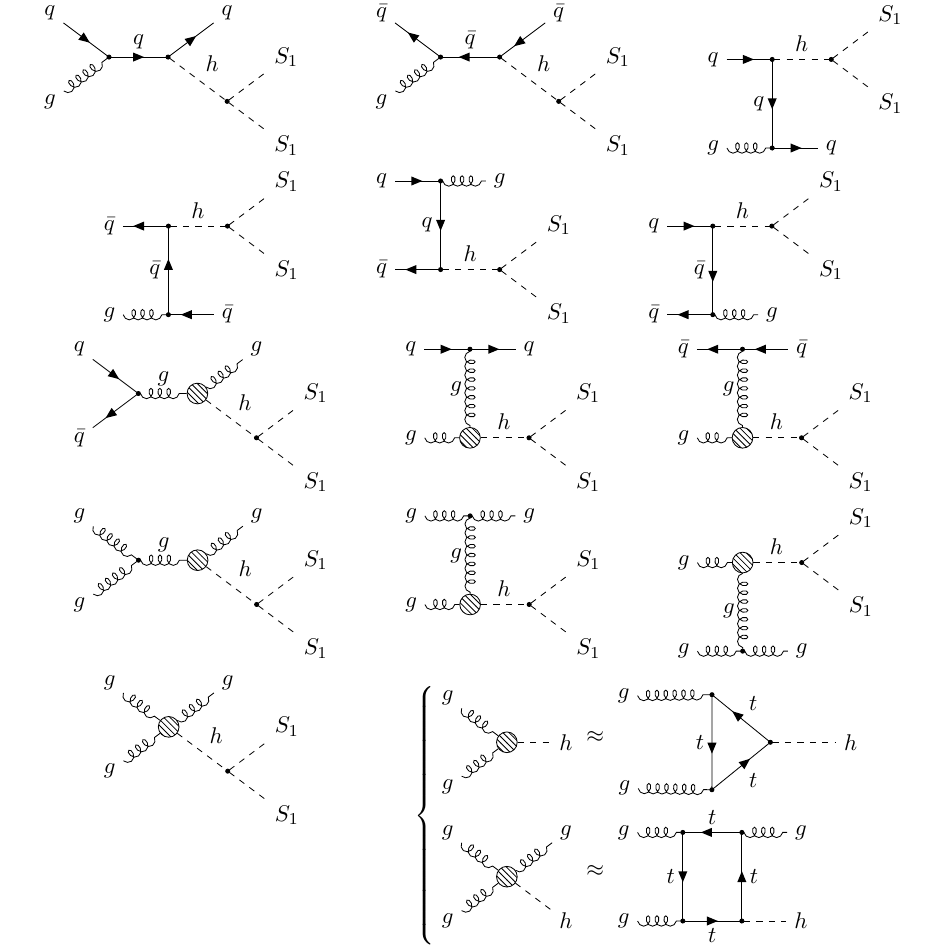}
    \caption{Feynman diagrams contributing to mono-jet production processes $p p \rightarrow S_1 S_1 \text{j}$ at tree-level, where $\text{j}=q,\bar{q}, g$ are jets and $q=u,d,c,s,b$. The vertices for the $ggh$ and $gggh$ effective couplings arise from top quark loops (in gluon fusion) in the heavy top quark limit ($m_t^2 \gg s$).}
    \label{fig:pp->S1S1j diagrams}
\end{figure}
\begin{figure}[!ht]
    \centering
    \includegraphics[width=0.95\textwidth]{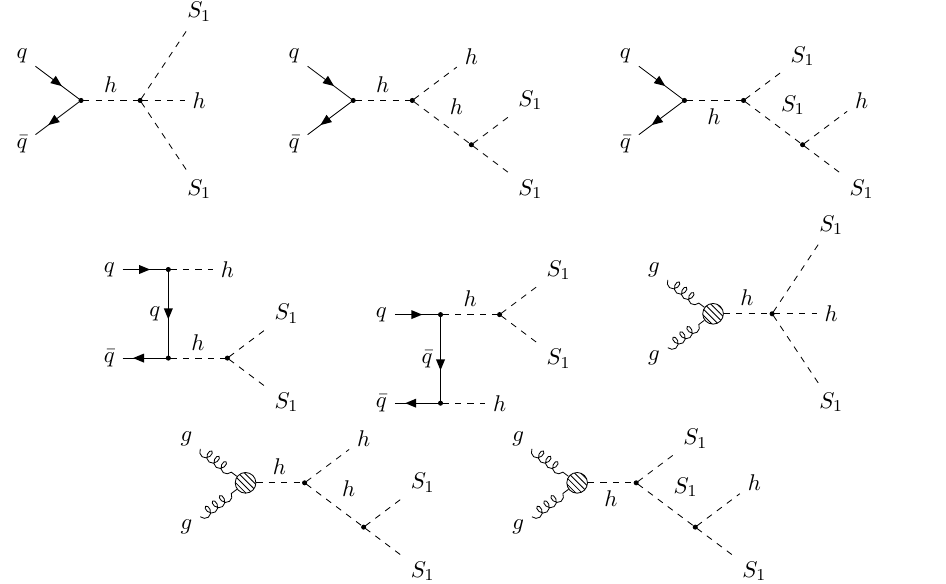}
    \caption{Feynman diagrams contributing to the mono-Higgs production process $p p \rightarrow S_1 S_1 h$ at tree-level, where $q=u,d,c,s,b$. The $ggh$ effective vertex is explicitly shown in Fig.~\ref{fig:pp->S1S1j diagrams}.}
    \label{fig:pp->S1S1h diagrams}
\end{figure}
\begin{figure}[!ht]
    \centering
    \includegraphics[width=0.95\textwidth]{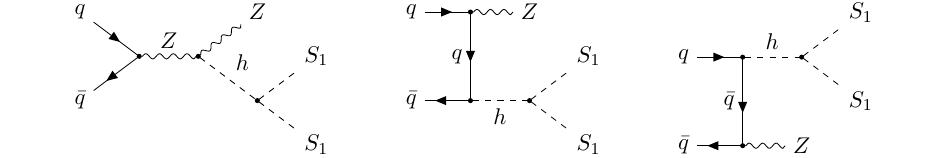}
    \caption{Feynman diagrams contributing to the mono-$Z$ production process $p p \rightarrow S_1 S_1 Z$ at tree-level, where $q=u,d,c,s,b$.}
    \label{fig:pp->S1S1Z diagrams}
\end{figure}
%

In the model with just one singlet the DM production cross sections are negligible because DM masses are above about 3.5 TeV. In these multi-singlet extensions, however, the production cross sections are of the order of any other weak process at the LHC. Therefore, one may ask if these light DM states can be probed at the present or future runs. Since DM particles cannot be directly detected at the LHC (or any other collider), we must consider processes that in addition include visible particles in the final state. The collider events associated with such processes have a large amount of missing energy.  At the LHC, these processes are $pp \rightarrow X + \slashed{E}_{\text{T}}$, where $X$ and $\slashed{E}_{\text{T}} \equiv E_{\text{T}}^{\text{miss}}$ stand for SM visible particles and \emph{missing transverse energy} (MET), respectively.

\begin{table}[!ht]
\centering
\begin{tabular}{|c|c|cc|cc|cc|}
\hline
\multirow{2}{*}{$m_{S_1}$ [GeV]} & \multirow{2}{*}{$\kappa_{H1}$}  & \multicolumn{2}{c|}{$\sqrt{s}=13$ TeV} & \multicolumn{2}{c|}{$\sqrt{s}=13.6$ TeV} & \multicolumn{2}{c|}{$\sqrt{s}=14$ TeV} \\
\cline{3-8}
& & $\sigma$ [fb] & NE & $\sigma$ [fb] & NE & $\sigma$ [fb] & NE \\
\hline
124.8 & 8.821 & 346.800 & 65892 & 386.000 & 173700 & 412.100 & 1236300 \\
140.0 & 5.700 & 82.800 & 15732 & 92.320 & 41544 & 99.090 & 297270 \\
155.0 & 8.398 & 110.400 & 20976 & 123.500 & 55575 & 132.800 & 398400 \\
170.4 & 8.191 & 66.460 & 12627 & 74.690 & 33610 & 80.310 & 240930 \\
185.3 & 8.350 & 46.310 & 8799 & 52.070 & 23431 & 56.190 & 168570 \\
201.9 & 9.548 & 40.030 & 7606 & 45.190 & 20335 & 48.840 & 146520 \\
\hline
\end{tabular}
\caption{Cross section $\sigma(p p \rightarrow S_1 S_1 \text{j})$ and corresponding number of events (NE) for six benchmark points from the two singlet model, for three center-of-mass energies. 
The NE's were computed for the following integrated luminosities: $\text{L}= 190 \text{ fb}^{-1}$ at $\sqrt{s}=13$ TeV, $\text{L}= 450 \text{ fb}^{-1}$ at $\sqrt{s}=13.6$ TeV and $\text{L}= 3000 \text{ fb}^{-1}$ at $\sqrt{s}=14$ TeV.}
\label{tab:mono-jet}
\end{table}
\begin{table}[!ht]
\centering
\begin{tabular}{|c|c|cc|cc|cc|}
\hline
\multirow{2}{*}{$m_{S_1}$ [GeV]} & \multirow{2}{*}{$\kappa_{H1}$}  & \multicolumn{2}{c|}{$\sqrt{s}=13$ TeV} & \multicolumn{2}{c|}{$\sqrt{s}=13.6$ TeV} & \multicolumn{2}{c|}{$\sqrt{s}=14$ TeV} \\
\cline{3-8}
& & $\sigma$ [fb] & NE & $\sigma$ [fb] & NE & $\sigma$ [fb] & NE \\
\hline
124.8 & 8.821 & 18.500 & 3515 & 20.760 & 9342 & 22.340 & 67020 \\
140.0 & 5.700 & 2.337 & 444 & 2.629 & 1183 & 2.835 & 8505  \\
155.0 & 8.398 & 6.477 & 1231 & 7.318 & 3293 & 7.907 & 23721 \\
170.4 & 8.191 & 3.958 & 752 & 4.477 & 2015 & 4.843 & 14529 \\
185.3 & 8.350 & 2.950 & 560 & 3.348 & 1507 & 3.628 & 10884 \\
201.9 & 9.548 & 3.317 & 630 & 3.774 & 1698 & 4.093 & 12279 \\
\hline
\end{tabular}
\caption{Cross section $\sigma(p p \rightarrow S_1 S_1 h)$ and corresponding number of events (NE) for six benchmark points from the two singlet model, for three center-of-mass energies. The NE's were computed for the following integrated luminosities: $\text{L}= 190 \text{ fb}^{-1}$ at $\sqrt{s}=13$ TeV, $\text{L}= 450 \text{ fb}^{-1}$ at $\sqrt{s}=13.6$ TeV and $\text{L}= 3000 \text{ fb}^{-1}$ at $\sqrt{s}=14$ TeV.}
\label{tab:mono-h}
\end{table}
\begin{table}[!ht]
\centering
\begin{tabular}{|c|c|cc|cc|cc|}
\hline
\multirow{2}{*}{$m_{S_1}$ [GeV]} & \multirow{2}{*}{$\kappa_{H1}$}  & \multicolumn{2}{c|}{$\sqrt{s}=13$ TeV} & \multicolumn{2}{c|}{$\sqrt{s}=13.6$ TeV} & \multicolumn{2}{c|}{$\sqrt{s}=14$ TeV} \\
\cline{3-8}
& & $\sigma$ [fb] & NE & $\sigma$ [fb] & NE & $\sigma$ [fb] & NE \\
\hline
124.8 & 8.821 & 2.887 & 548 & 3.119 & 1403 & 3.270 & 1471 \\
140.0 & 5.700 & 0.557 & 106 & 0.603 & 271 & 0.634 & 285 \\
155.0 & 8.398 & 0.615 & 117 & 0.672 & 302 & 0.705 & 317 \\
170.4 & 8.191 & 0.316 & 60 & 0.345 & 155 & 0.364 & 164 \\
185.3 & 8.350 & 0.190 & 36 & 0.208 & 93 & 0.220 & 99 \\
201.9 & 9.548 & 0.140 & 27 & 0.154 & 69 & 0.163 & 73 \\
\hline
\end{tabular}
\caption{Cross section $\sigma(p p \rightarrow S_1 S_1 Z)$ and corresponding number of events (NE) for six benchmark points from the two singlet model, for three center-of-mass energies. The NE's were computed for the following integrated luminosities: $\text{L}= 190 \text{ fb}^{-1}$ at $\sqrt{s}=13$ TeV, $\text{L}= 450 \text{ fb}^{-1}$ at $\sqrt{s}=13.6$ TeV and $\text{L}= 3000 \text{ fb}^{-1}$ at $\sqrt{s}=14$ TeV.}
\label{tab:mono-Z}
\end{table}
\begin{table}[!ht]
\centering
\begin{tabular}{|c|c|cc|cc|cc|}
\hline
\multirow{2}{*}{$m_{S_1}$ [GeV]} & \multirow{2}{*}{$\kappa_{H1}$}  & \multicolumn{2}{c|}{$\sqrt{s}=13$ TeV} & \multicolumn{2}{c|}{$\sqrt{s}=13.6$ TeV} & \multicolumn{2}{c|}{$\sqrt{s}=14$ TeV} \\
\cline{3-8}
& & $\sigma$ [fb] & NE & $\sigma$ [fb] & NE & $\sigma$ [fb] & NE \\
\hline
124.8 & 8.821 & 31.180 & 5924 & 34.940 & 15723 & 37.510 & 112530 \\
140.0 & 5.700 & 4.834 & 918 & 5.414 & 2436 & 5.810 & 17430 \\
155.0 & 8.398 & 9.951 & 1891 & 11.210 & 5044 & 12.070 & 36210 \\
170.4 & 8.191 & 5.911 & 1123 & 6.672 & 3002 & 7.216 & 21648 \\
185.3 & 8.350 & 4.258 & 809 & 4.825 & 2171 & 5.214 & 15642 \\
201.9 & 9.548 & 4.474 & 850 & 5.086 & 2289 & 5.512 & 16536 \\
\hline
\end{tabular}
\caption{Cross section $\sigma(p p \rightarrow S_1 S_1 b \bar{b})$ and corresponding number of events (NE) for six benchmark points from the two singlet model, for three center-of-mass energies. The NE's were computed for the following integrated luminosities: $\text{L}= 190 \text{ fb}^{-1}$ at $\sqrt{s}=13$ TeV, $\text{L}= 450 \text{ fb}^{-1}$ at $\sqrt{s}=13.6$ TeV and $\text{L}= 3000 \text{ fb}^{-1}$ at $\sqrt{s}=14$ TeV.}
\label{tab:bb_bar}
\end{table}

Let us now discuss the most common processes for which there are already experimental analyses by ATLAS and/or CMS. We start by computing the cross sections for the so-called mono-$X$ processes (where $X$ is a single SM particle), for six allowed points in the two singlets scenario, with \texttt{MadGraph5\_aMC@NLO}~\cite{Alwall:2011uj, Alwall:2014hca}. The computations were performed at LO for center-of-mass energies ($\sqrt{s}$) of 13 TeV, 13.6 TeV and 14 TeV, with corresponding integrated luminosities (L) of 190 fb$^{-1}$, 450 fb$^{-1}$ and 3000 fb$^{-1}$, corresponding to the different LHC runs including the High-Luminosity (HL-LHC) stage \cite{Zerlauth:2024aeh}. We have considered the processes
\begin{equation}
	p p \rightarrow S_1 S_1 X \, , \qquad X=\text{j}, h, Z \, ,
\end{equation}
where $\text{j}=q, \bar{q}, g$ are jets and $q=u,d,c,s,b$. The corresponding tree-level Feynman diagrams are presented in Figs.~\ref{fig:pp->S1S1j diagrams},~\ref{fig:pp->S1S1h diagrams} and~\ref{fig:pp->S1S1Z diagrams}, respectively. Additionally, we also considered the di-jet process of $p p \rightarrow S_1 S_1 b \bar{b}$.
The results for the mono-jet, mono-Higgs, mono-$Z$ and di-jet $b \bar{b}$ pair production are presented in Tables~\ref{tab:mono-jet},~\ref{tab:mono-h},~\ref{tab:mono-Z} and~\ref{tab:bb_bar}, respectively. The Feynman diagrams containing $ggh$ effective vertices -- which arise from top quark loops in gluon fusion (see Fig.~\ref{fig:pp->S1S1j diagrams}) -- are the main contributors to the cross sections of the mono-jet, mono-Higgs and di-jet processes. On the other hand, all Feynman diagrams for the mono-$Z$ process also feature a Higgs mediator, but (at tree-level) do not contain the dominant $ggh$ effective vertex. Hence, the mono-$Z$ cross sections (and number of events) are lower than for the previous processes, as shown in the tables.
%
In the next sections we will explore all these final states in more detail, taking into account LHC searches already performed. To this end, we focus on the ATLAS analyses~\cite{ATLAS:2021kxv,ATLAS:2021shl,ATLAS:2018nda}, which provide the most recent (to the best of our knowledge) model-independent upper limits. Other relevant searches also exist, such as the similar analyses~\cite{CMS:2021far,CMS:2019ykj,CMS:2020ulv} by the CMS Collaboration.

\subsection{Jet Searches}
\label{sec:SM+2RSS mono-jet searches}

In Ref.~\cite{ATLAS:2021kxv}, the ATLAS Collaboration established model-independent upper limits at the 95\% confidence level (CL) on the visible cross section of DM production processes with one to four (final state) jets $\text{j} = q, \bar{q}, g$, where $q = u,d,c,s,b$, at a center-of-mass energy of $\sqrt{s}=13$ TeV. 
This visible cross section is defined as
\begin{equation}
	\sum_{N=1}^{4} \sigma^{\text{vis}}(pp \rightarrow N \text{j} + \text{DM}) \equiv \sum_{N=1}^{4} \sigma(pp \rightarrow N \text{j} + \text{DM}) \times A \times \epsilon \, , \label{eq:visible cross section jets}
\end{equation}
where $A$ is the kinematic acceptance and $\epsilon$ is the experimental efficiency. 
These upper limits are set as a function of the \emph{cut} on the missing transverse momentum, so that $p_{\text{T}}^{\text{miss}}$ is larger than the considered $p_{\text{T}}^{\text{miss}}$ cut (or lower bound).
The acceptances $A$ and efficiencies $\epsilon$ were not provided and we took them equal to $1$ because the model cross sections are still more than one order of magnitude away from being probed, as we will see later.

These limits were obtained from an analysis of data recorded by the ATLAS detector at the LHC during the 2015-2018 time period of Run 2, corresponding to an integrated luminosity of $\text{L}=139$ fb$^{-1}$ at $\sqrt{s} = 13$ TeV.  They are part of a broader search for beyond-SM (BSM) new physics phenomena in events with a final state containing one to four jets and large missing transverse momentum in proton-proton collisions. The final states of selected events for this BSM phenomena search~\cite{ATLAS:2021kxv} were required to meet the following criteria:
\begin{itemize}
	\item large missing transverse momentum, $p_{\text{T}}^{\text{miss}}>200$ GeV;
	\item a leading jet with transverse momentum $p_{\text{T}}>150$ GeV and pseudorapidity $|\eta|<2.4$, along with none to three additional (non-leading) jets with transverse momenta $p_{\text{T}}>30$ GeV and pseudorapidities $|\eta|<2.8$;
	\item azimuthal angle separations $\Delta \phi (\vec{p}^{\, \text{miss}}_{\text{T}}, \vec{p}^{ \, \text{j}_i}) \equiv \phi (\vec{p}^{\, \text{miss}}_{\text{T}}) - \phi (\vec{p}^{ \, \text{j}_i})$ between the missing transverse 3-momentum $\vec{p}^{\, \text{miss}}_{\text{T}}$ and the 3-momentum $\vec{p}^{ \, \text{j}_i}$ of each jet $\text{j}_i$, $i=1, \dots, N$ (with $N \in \{ 1,2,3,4 \}$), satisfying
	\begin{equation}
		\Delta \phi (\vec{p}^{\, \text{miss}}_{\text{T}}, \vec{p}^{ \, \text{j}_i}) >
		\begin{cases}
			0.4 \text{ rad} \, , \quad p^{\, \text{miss}}_{\text{T}} > 250 \text{ GeV}\\
			0.6 \text{ rad} \, , \quad 200 \text{ GeV} < p^{\, \text{miss}}_{\text{T}} \leq 250 \text{ GeV}
		\end{cases} \, .
	\end{equation}
\end{itemize}

Using \texttt{MadGraph5\_aMC@NLO}~\cite{Alwall:2011uj, Alwall:2014hca}, we computed the cross section $\sum_{N=1}^{4} \sigma(pp \rightarrow S_1 S_1 + N \text{j})$ of the $S_1$ pair plus one to four jet production processes at $\sqrt{s}=13$ TeV.

\begin{figure}[!ht]
    \centering
    \includegraphics[width=\textwidth]{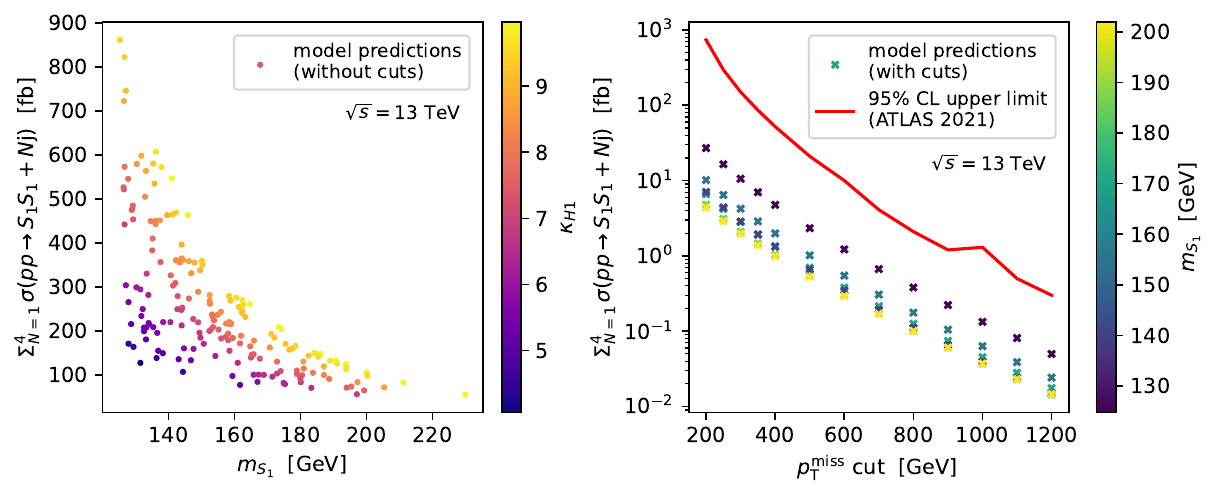}
    \caption{Cross section of jet production processes $p p \rightarrow S_1 S_1+N \text{j}$ ($N=1,2,3,4$), where $\text{j}=q,\bar{q},g$ and $q=u,d,c,s,b$, at a center-of-mass energy of $\sqrt{s}=13$ TeV. The red (solid) line is the ATLAS (2021) experimental upper limit on the visible cross section ($\sigma^{\text{vis}}$) of $pp \rightarrow N \text{j}+\text{DM}$  ($N=1,2,3,4$) processes. Left panel: cross section with no cuts. Right panel: the six benchmark points with several missing transverse momentum ($p_{\text{T}}^{\text{miss}}$) cuts, along with the corresponding ATLAS upper limits. The colour bar in the left (right) plot represents the portal coupling (the mass) of $S_1$. The benchmark points can be identified via their colour: the DM mass increases going from violet to yellow from $124.8~$GeV to $201.9~$GeV.}
    \label{fig:SM+2RSS mono-jet}
\end{figure}

The results are shown in Fig.~\ref{fig:SM+2RSS mono-jet} displaying the cross sections $\sum_{N=1}^{4} \sigma(pp \rightarrow S_1 S_1 + N \text{j})$ represented as dots (left panel) or crosses (right panel) -- and the ATLAS upper limit on the visible cross section~Eq.~(\ref{eq:visible cross section jets}) for several missing transverse momentum cuts, represented by a solid red line (right panel). The left panel shows the cross section with no cuts plotted as a function of $m_{S_1}$ and the portal coupling constant $\kappa_{H1}$ in the colour bar. The right panel presents the ATLAS upper limits on the visible cross section~Eq.~(\ref{eq:visible cross section jets}) as a function of the missing transverse momentum cut for the six benchmark points after applying the previously presented ATLAS selection criteria. One can easily conclude that all points are slightly more than one order of magnitude away from being experimentally probed by this LHC search.

\subsection{Mono-Higgs Searches}
\label{sec:SM+2RSS mono-higgs searches}

In Ref.~\cite{ATLAS:2021shl}, the ATLAS Collaboration established model-independent upper limits at the 95\% CL on the visible cross section of DM production processes with a (final state) Higgs boson, at a center-of-mass energy of $\sqrt{s}=13$ TeV. This visible cross section is defined as
\begin{equation}
	\sigma^{\text{vis}}(pp \rightarrow h + \text{DM}) \equiv \sigma(pp \rightarrow h + \text{DM}) \times A \times \epsilon \times \text{BR}(h \rightarrow b \bar{b}) \, , \label{eq:visible cross section mono-higgs}
\end{equation}
where $A$ is the acceptance, $\epsilon$ is the efficiency and the branching ratio for the hadronic Higgs boson decay is (defined as, and) given by $\text{BR}(h \rightarrow b \bar{b}) \equiv \Gamma(h \rightarrow b \bar{b}) / \Gamma_h^{\text{(total)}} \approx 0.58$. Since all three factors are positive and smaller than one, the visible cross section is lower than the cross section, i.e., $\sigma^{\text{vis}} \equiv \sigma \times A \times \epsilon \times \text{BR}(h \rightarrow b \bar{b}) < \sigma$. These upper limits are set as a function of the missing transverse momentum \emph{range}, where $p_{\text{T}}^{\text{miss}}$ is both bounded from below and from above.

These limits were obtained from an analysis of data recorded by the ATLAS detector at the LHC during Run 2 -- corresponding to an integrated luminosity of $\text{L}=139$ fb$^{-1}$ at a center-of-mass energy of $\sqrt{s} = 13$ TeV -- and are part of a broader search for DM particles produced in mono-Higgs events. The event selection can be consulted in (section 5 of) Ref.~\cite{ATLAS:2021shl}.

\begin{figure}[!th]
    \centering
    \includegraphics[width=\textwidth]{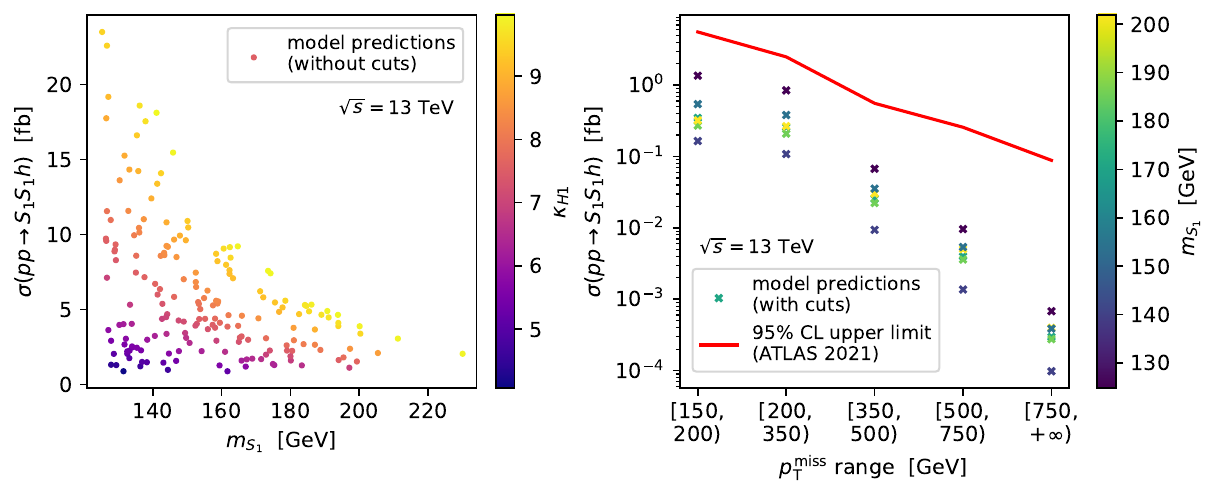}
    \caption{Cross section of mono-Higgs production processes $p p \rightarrow S_1 S_1 h$ at a center-of-mass energy of $\sqrt{s}=13$ TeV for the chosen benchmark points. The red line is the ATLAS (2021) model-independent experimental upper limit on the $\sigma^{\text{vis}}/\text{BR}(h \to b \bar{b}) \equiv \sigma \times A \times \epsilon$ of $pp \rightarrow Z+\text{DM}$ process. Left panel: model predictions for the chosen scenarios. Right panel: cross section for the chosen six benchmark points with different missing transverse momentum ($p_{\text{T}}^{\text{miss}}$) ranges, along with the corresponding ATLAS upper limits. The colour bar in the left (right) plot represents
the portal coupling (the mass) of $S_1$.}
    \label{fig:SM+2RSS mono-higgs}
\end{figure}

Using \texttt{MadGraph5\_aMC@NLO}~\cite{Alwall:2011uj, Alwall:2014hca}, we computed the cross section $\sigma(pp \rightarrow S_1 S_1 h)$ of the $S_1$ pair plus mono-Higgs production process at $\sqrt{s}=13$ TeV. Analogously to Sec.~\ref{sec:SM+2RSS mono-jet searches} for the jet search analysis~\cite{ATLAS:2021kxv}, this mono-Higgs search analysis~\cite{ATLAS:2021shl} does not provide the acceptances $A$ or efficiencies $\epsilon$ corresponding to the ATLAS upper limits on the visible cross section. 
We take them equal to $1$ because they are still close to one order of magnitude away from being probed as can be seen in Fig.~\ref{fig:SM+2RSS mono-higgs} (right panel). The figure shows the cross section $\sigma(pp \rightarrow S_1 S_1 h)$ for the model -- represented as dots (left panel) or crosses (right panel) -- and the ATLAS upper limit on $\sigma^{\text{vis}}/\text{BR}(h \to b \bar{b}) \equiv \sigma \times A \times \epsilon$ ($<\sigma$) for distinct missing transverse momentum ranges, represented by a red line. The left panel shows the model predictions for the cross section without any momentum cuts plotted as a function of the mass $m_{S_1}$ and the portal coupling constant $\kappa_{H1}$ (colour bar) of $S_1$. The right panel presents the ATLAS upper limits on $\sigma^{\text{vis}}/\text{BR}(h \to b \bar{b}) \equiv \sigma \times A \times \epsilon$ as a function of the missing transverse momentum range, alongside the model predictions for the chosen six benchmark points with the same cuts.  One concludes that the points are not excluded by the current LHC results by slightly less than one order of magnitude.
A more detailed analysis has to be performed once more data is gathered.

\subsection{Mono-$Z$ Searches}
\label{sec:SM+2RSS mono-Z searches}

In Ref.~\cite{ATLAS:2018nda}, the ATLAS Collaboration established model-independent upper limits at the 95\% CL on the visible cross section of DM production processes with a $Z$ boson in the final state, at $\sqrt{s}=13$ TeV. This visible cross section is defined as
\begin{equation}
	\sigma^{\text{vis}} (pp \rightarrow Z + \text{DM}) \equiv \sigma(pp \rightarrow Z + \text{DM}) \times A \times \epsilon \times \sum_{q} \text{BR}(Z \rightarrow q \bar{q}) \, , \label{eq:visible cross section mono-Z}
\end{equation}
where $A$ is the acceptance, $\epsilon$ is the efficiency and BR the branching ratio for the hadronic $Z$ boson decay ($\approx 0.69$). As for the jets, the upper limits are set as a function of the missing transverse momentum \emph{range}.  These limits were obtained from an analysis of data recorded by the ATLAS detector at the LHC during the 2015-2016 time period of Run 2,  corresponding to an integrated luminosity of $\text{L}=36.1$ fb$^{-1}$ at $\sqrt{s} = 13$ TeV. They are part of a broader search for DM particles produced in mono-$W^\pm$/$Z$ events. The event selection details can be found in~\cite{ATLAS:2018nda}.

\begin{figure}[!ht]
    \centering
    \includegraphics[width=\textwidth]{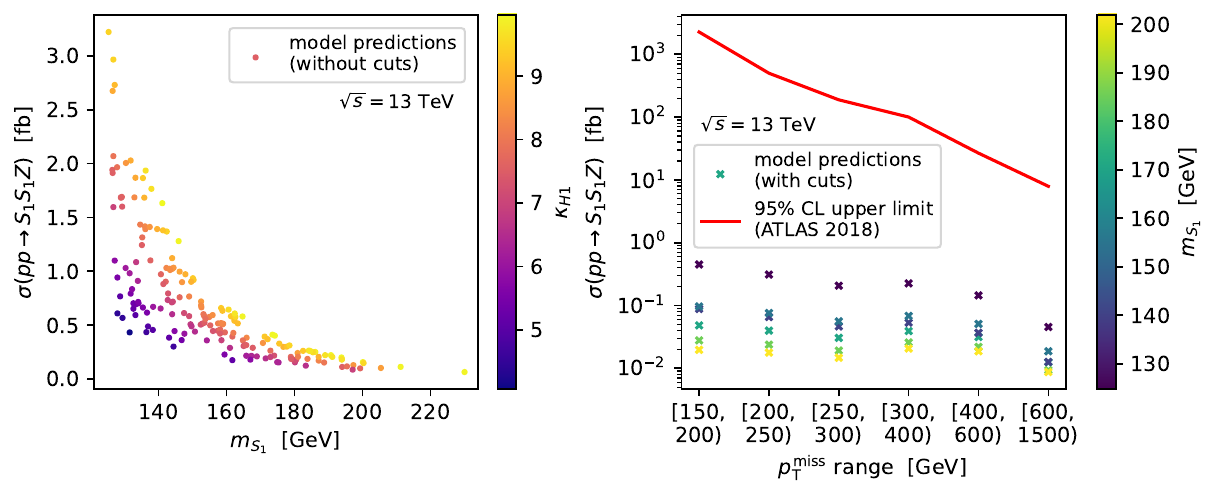}
    \caption{Cross section for $p p \rightarrow S_1 S_1 Z$ for $\sqrt{s}=13$ TeV. The red (solid) line is the ATLAS (2018)  experimental upper limit on the cross section ($\sigma$) of $pp \rightarrow Z+\text{DM}$. Left panel: cross section with no cuts. Right panel: benchmark points for several missing transverse momentum ($p_{\text{T}}^{\text{miss}}$) ranges, along with the corresponding ATLAS upper limits. The colour bar in the left (right) plot represents the portal coupling (the mass) of $S_1$.}
    \label{fig:SM+2RSS mono-Z}
\end{figure}

Using \texttt{MadGraph5\_aMC@NLO}~\cite{Alwall:2011uj, Alwall:2014hca}, we computed the cross section $\sigma(pp \rightarrow S_1 S_1 Z)$ of the $S_1$ pair plus $Z$ production process at a center-of-mass energy of $\sqrt{s}=13$ TeV. In this search~\cite{ATLAS:2018nda} both the acceptances $A$ and the efficiencies $\epsilon$ are provided for the upper limits on the cross section as a function of the missing transverse momentum. 

The results are shown in Fig.~\ref{fig:SM+2RSS mono-Z},  dots (left panel) or crosses (right panel), and the ATLAS upper limit on the cross section, Eq.~\eqref{eq:visible cross section mono-Z}, for several missing transverse momentum ranges, represented by a solid red line (right panel). The left panel shows the cross section with no cuts as a function of the mass $m_{S_1}$ and with the portal coupling constant $\kappa_{H1}$ (colour bar) of $S_1$. The right panel presents the ATLAS upper limits on the cross section as a function of the missing transverse momentum range, with the benchmark points with the same momentum cuts. One can easily conclude that all points are far from being experimentally probed by this LHC search. In particular, the right panel alone shows that these points are roughly three to five orders of magnitude away from exclusion. 

In this section we have discussed scalar DM searches at the LHC. The most promising channels are the ones with Higgs and jets in the final state. We believe there are good chances to probe these models at the High-Luminosity stage of the LHC.

\section{Conclusions \label{sec:conclusions}}

We studied real scalar singlet extensions of the SM with unbroken discrete symmetries, which may give origin to several DM candidates. In these models the SM is unaltered in all interactions between SM particles. There are, however, one or more portal terms that connect the Higgs field with each of the dark sectors. Our starting point was to add just one real scalar singlet to the SM, a well-known DM model that had already been extensively studied. We have shown that with the most recent constraints the model is experimentally excluded for DM masses below about 3500 GeV.  This makes the model impossible to probe at the LHC. The other allowed region is for a DM mass equal to half of the Higgs mass. In this scenario, resonant contributions to the relic density allow for a smaller portal coupling which in turn permits DD bounds to be evaded.

We then proceeded to study the two real scalar singlets extension of the SM. In this model we again  found the allowed regions of heavy masses and the one where one of the DM candidates has half the Higgs mass. However,  a new allowed region was uncovered where one of the DM particles is light, just above the Higgs mass, and the other one is heavy.  This new scenario is only possible in the two singlets case due to the fact that the two DM particles share the relic density. The heavier DM particle is responsible for almost all the observed DM relic density which means that its mass and portal coupling are highly constrained due to the tension between relic density and DD exclusion. In contrast, the lighter DM particle has a negligible contribution of $\Omega_{S_1}h^2 \sim 10^{-7}$ to the total DM relic density, meaning that  the DD cross section of interest is now given by $\sigma^{\text{SI}}(S_1 N \rightarrow S_1 N) \times \Omega_{S_1}/\Omega_{\text{DM}}$. 
We also concluded that the inter-dark coefficient $\lambda_{12}$ regulating the heavy to light $S_2 S_2 \rightarrow S_1 S_1$ annihilation does not influence the total DM relic density as significantly as $m_{S_2}$ and $\kappa_{H2}$, and is therefore not constrained by it. But since it affects the $S_1$ fraction $\Omega_{S_1}/\Omega_{\text{DM}}$, it is bounded from above due to DD exclusion related to $S_1$. This new allowed region turned out to be almost excluded by the new LUX-ZEPLIN results of 2024 with most points falling inside the experimental uncertainty band for both particles.

We then moved to the model with two singlets but with only one dark symmetry. In this scenario the lightest scalar from the dark sector is now allowed to be in the entire region above the Higgs mass. Due to the existence of two scalars in the dark sector and only one symmetry, the rotation to mass eigenstates redefines the couplings in such a way that only one of the portal couplings is related to DD while all three portal couplings contribute to the relic density.
Therefore, the couplings can be tuned to be in agreement with all experimental results.

In order to see if new regions would open up in the case of three singlets with three different dark symmetries we also studied this model. Besides the scenarios already expected we could now have one light and two heavy and two light and one heavy scalars. The latter is very similar to the case with two singlets. The former has the new feature of having two heavy DM particles simultaneously accounting for the observed relic density. Consequently, the DD exclusion related to the heavy DM particles is now determined also by their abundance fractions. Therefore, the tension between relic density and DD exclusion decreases, slightly weakening the usually stringent constraints on the heavy DM particle masses and portal couplings. Adding more singlets would make the bounds on masses and portal couplings  increasingly loose.

Finally, we selected the most relevant LHC experiments to determine whether our allowed points were experimentally excluded by collider searches. Note that our benchmark points have very large portal couplings. Experiments include searches performed with the ATLAS and CMS detectors of events of up to four jets and missing energy, mono-Higgs and mono-$Z$ events. The searches established upper limits on the cross section for the corresponding DM production processes. In all  the models presented in this work, that is, singlet extensions of the SM with DM candidates,  the cross sections are close to being probed
in case of the mono-Higgs searches, slightly above one order of magnitude away for events with jets in the final state and orders of magnitude away from being excluded in the case of mono-$Z$. This suggests that detecting a scalar WIMP at the LHC in this type of models could happen in the High-Luminosity stage.

\subsubsection*{Acknowledgments}
RS, MG and TT are are partially supported by the Portuguese Foundation for Science and Technology (FCT) under CFTC: UIDB/00618/2020 (\url{https://doi.org/10.54499/UIDB/00618/2020}), UIDP/00618/2020
(\url{https://doi.org/10.54499/UIDP/00618/2020}) and through the PRR (Recovery and Resilience
Plan), within the scope of the investment “RE-C06-i06 - Science Plus Capacity Building”, measure “RE-C06-i06.m02 - Reinforcement of financing for International Partnerships in Science,
Technology and Innovation of the PRR”, under the project with the reference 2024.03328.CERN (\url{https://doi.org/10.54499/2024.03328.CERN}). MG is additionally supported by FCT with a PhD Grant (reference 2023.02783.BD). The work of M.M.\ is supported by the BMBF-Project 05H24VKB.
We thank Guglielmo Frattari and  Valerio Ippolito for discussions.

\vspace*{1cm}
\bibliographystyle{h-physrev}
\bibliography{references.bib}

\end{document}